\documentclass[aps,pre,twocolumn,superscriptaddress,floatfix]{revtex4-1}
\usepackage[usenames,dvipsnames]{color}
\usepackage{graphicx} % Allows for eps images
\usepackage{epstopdf}
\usepackage{epsfig}
\usepackage{amsmath} % AMS Math Package
\usepackage{amsthm} % Theorem Formatting
\usepackage{amssymb}	% Math symbols such as \mathbb

\usepackage{hyperref} %clickable references.
\hypersetup{
    colorlinks,
    citecolor=blue,
    filecolor=black,
    linkcolor=blue,
    urlcolor=blue
}

\usepackage{tikz}
\usetikzlibrary{patterns}

\newcommand{\eq}[1]{\begin{align} #1 \end{align}}
\newcommand{\eqs}[1]{\begin{align} \begin{split} #1 \end{split}\end{align}}
\newcommand{\p}[0]{^\prime}

\newcommand{\Nc}[0]{\ensuremath{N_\text{c}}}
\newcommand{\Nciso}[0]{\ensuremath{N_\text{c}^\text{iso}}}
\newcommand{\Nbndry}[0]{\ensuremath{N_\text{dof}^\text{bndry}}}

\newcommand{\Nzm}[0]{\ensuremath{N_\text{zm}}}

\newcommand{\He}[0]{\ensuremath{\hat K}}
\newcommand{\Ziso}[0]{\ensuremath{Z_\text{iso}^N}}
\newcommand{\Zmin}[0]{\ensuremath{Z^N_\text{min}}}
\newcommand{\Gdc}[0]{\ensuremath{G_{DC}}}

\newcommand{\Rc}[0]{\ensuremath{\mathcal R_{comp}}}
\newcommand{\Ra}[0]{\ensuremath{\mathcal R_{all}}}
\newcommand{\Rap}[0]{\ensuremath{\mathcal R_{all}^+}}
\newcommand{\Ec}[0]{\ensuremath{\mathcal E_{comp}}}
\newcommand{\Ea}[0]{\ensuremath{\mathcal E_{all}}}
\newcommand{\Eap}[0]{\ensuremath{\mathcal E_{all}^+}}

\renewcommand{\v}[1]{\ensuremath{\mathbf{#1}}} % for vectors

\newcommand{\avg}[1]{\left< #1 \right>} % for average

\begin{document}

% Use the \preprint command to place your local institutional report
% number in the upper righthand corner of the title page in preprint mode.
% Multiple \preprint commands are allowed.
% Use the 'preprintnumbers' class option to override journal defaults
% to display numbers if necessary
%\preprint{}

%Title of paper
\title{Jamming in finite systems: stability, anisotropy, fluctuations and scaling}

% repeat the \author .. \affiliation  etc. as needed
% \email, \thanks, \homepage, \altaffiliation all apply to the current
% author. Explanatory text should go in the []'s, actual e-mail
% address or url should go in the {}'s for \email and \homepage.
% Please use the appropriate macro foreach each type of information

% \affiliation command applies to all authors since the last
% \affiliation command. The \affiliation command should follow the
% other information
% \affiliation can be followed by \email, \homepage, \thanks as well.
\author{Carl P. Goodrich}
\email[]{cpgoodri@sas.upenn.edu}
\affiliation{Department of Physics, University of Pennsylvania, Philadelphia, Pennsylvania 19104, USA}

\author{Simon Dagois-Bohy}
\affiliation{Huygens-Kamerlingh Onnes Lab, Universiteit Leiden, Postbus 9504, 2300 RA Leiden, The Netherlands}
\affiliation{Instituut-Lorentz, Universiteit Leiden, Postbus 9506, 2300 RA Leiden, The Netherlands}

\author{Brian P. Tighe}
\affiliation{Delft University of Technology, Process \& Energy Laboratory, Leeghwaterstraat 39, 2628 CB Delft, The Netherlands} 

\author{Martin van Hecke}
\affiliation{Huygens-Kamerlingh Onnes Lab, Universiteit Leiden, Postbus 9504, 2300 RA Leiden, The Netherlands}

\author{Andrea J. Liu}
\affiliation{Department of Physics, University of Pennsylvania, Philadelphia, Pennsylvania 19104, USA}

\author{Sidney R. Nagel}
\affiliation{James Franck and Enrico Fermi Institutes, The University of Chicago, Chicago, Illinois 60637, USA}

%Collaboration name if desired (requires use of superscriptaddress
%option in \documentclass). \noaffiliation is required (may also be
%used with the \author command).
%\collaboration can be followed by \email, \homepage, \thanks as well.
%\collaboration{}
%\noaffiliation

\date{\today}

\begin{abstract}
Athermal packings of soft repulsive spheres exhibit a sharp jamming transition in the thermodynamic limit. Upon further compression, various structural and mechanical properties display clean power-law behavior over many decades in pressure. As with any phase transition, the rounding of such behavior in finite systems close to the transition plays an important role in understanding the nature of the transition itself. The situation for jamming is surprisingly rich: the assumption that jammed packings are isotropic is only strictly true in the large-size limit, and 
finite-size has a profound effect on the very meaning of jamming.
Here, we provide a comprehensive numerical study of finite-size effects in sphere packings above the jamming transition, focusing on stability as well as the scaling of the contact number and the elastic response.
\end{abstract}

% insert suggested PACS numbers in braces on next line
% insert suggested keywords - APS authors don't need to do this
%\keywords{}

%\maketitle must follow title, authors, abstract, \pacs, and \keywords
\maketitle
%\tableofcontents

\section{Introduction and Conclusions}

The theory of jammed amorphous solids has been largely based on packings at zero temperature of frictionless spheres with finite-range repulsions. Over the past decade, numerous studies have characterized the transition of such systems from an unjammed ``mechanical vacuum'' in which no particles interact at low packing fraction, $\phi$, to a jammed, rigid structure at high $\phi$ (see~\cite{Liu:2010jx,*vanHecke:2009go} and references therein). %\cite{OHern:2003vq,Liu:2010jx}\cite{grr}. 
The scenario that has emerged is that the jamming transition is a rare example of a random first-order transition%does this need a citation?
~\footnote{Note that the jamming transition appears to be a random first-order transition in dimensions $d \ge 2$, and is distinct from the glass transition, which is a random first-order transition in infinite dimensions~\cite{Parisi:2010uu}}. %In the large system limit, the bulk modulus $B$ and shear modulus $G$ both become non-zero for finite pressure --- 
%$G$ grows smoothly with pressure, whereas
%$B$ is discontinuous at zero pressure.
%Crucially, both moduli become non-zero simultaneously
%and concurrent with the onset of finite pressure --- thus making the transition both clear and well-defined.  
At the jamming transition, the average number of contacts per particle, $Z$, jumps discontinuously from zero to the value given by the rigidity criterion proposed originally by Maxwell.  Power-law scaling over many decades in confining pressure has been observed near the transition for the bulk modulus, shear modulus, energy, non-affinity, a characteristic frequency scale, various length scales 
and the excess contact number~\cite{Liu:2010jx,Durian:1995eo,OHern:2003vq,Silbert:2005vw,Wyart:2005vu,Wyart:2005jna,Wyart:2005wv,Silbert:2006bd,Ellenbroek:2006df,Ellenbroek:2009dp,Goodrich:2013ke}.
Moreover, the excess contact number and shear modulus have recently been shown to exhibit finite-size scaling, consistent with the critical nature of the jamming transition~\cite{Goodrich:2012ck}.

For ordinary critical phase transitions, singularities are rounded in finite systems but the nature of the transition remains qualitatively the same as it is in infinite ones.  
However, because the particle interactions in a jammed packing are purely repulsive and the force on every particle has to be balanced, a jammed packing must have a rigid structure that is system-spanning.  As a result, the nature of the boundary conditions is inextricably linked with the onset of rigidity, and boundary conditions play a particularly important role in finite jammed systems~\cite{Torquato:2001bm}.  For example, systems prepared in the standard way, in a fixed simulation box with periodic boundary conditions (that is, with the repeated zone of constant volume with {\em fixed} angles), can be unstable to shear even though they can support a pressure~\cite{DagoisBohy:2012dh}. 

Even for configurations that are stable to both shear and compression, the definition of the rigidity onset in terms of the development of nonzero bulk and shear moduli requires attention.  This is because jammed systems are only truly isotropic in the thermodynamic limit. Any finite system should properly be described by six elastic constants in 2 dimensions, or 21 in 3 dimensions, rather than the two elastic constants, the bulk and shear moduli, that describe isotropic systems.  Finally, the mechanical response of a finite system depends not only on the boundary conditions, but on whether or not the configuration has residual shear stress.  These considerations necessitate a careful reevaluation of jamming in finite systems. 

In this paper, we take all of these potential complications into account to develop a comprehensive finite-size analysis of compressed, athermal sphere packings with periodic-boundary conditions. 
We recast the $6$ ($21$) elastic constants needed in $2$ ($3$) dimensions in terms of (i) two combinations that are finite in the thermodynamic limit: the bulk modulus, $B$, and $G_{DC}$ (which approaches the shear modulus in the thermodynamic limit) and (ii) three combinations that measure anisotropic fluctuations and vanish in that limit.  Despite the complications alluded to above, for all of the ensembles studied and independent of the criteria used to identify the jamming transition, we show that $pN^2$ (where $p$ is the pressure and $N$ is the system size) is the correct scaling variable for the key quantities of excess contact number, $B$ and $G_{DC}$.  This is
consistent with earlier results for one of these ensembles~\cite{Goodrich:2012ck}.  
(In the case of two dimensions, our results are consistent with the presence of logarithmic corrections to scaling, supporting the conjecture~\cite{Liu:2010jx,Charbonneau:2012fl,Goodrich:2012ck,Wyart:2005vu} that the upper critical dimension for jamming is $d=2$.)

One of the three elastic constants that vanish in the thermodynamic limit also collapses with $pN^2$ and vanishs in the limit of $pN^2 \rightarrow \infty$ as $1/\sqrt{N}$. This is consistent with the central-limit theorem. The remaining two exhibit this behavior only for ensembles that have zero residual shear stress.  Thus, for the ensembles with no shear stress, we observe scaling collapse with $pN^2$ for all variables studied.

We note that one consequence of the scaling collapse with $pN^2$ is that one needs larger and larger systems as the jamming transition is approached to be in the thermodynamic limit. If the limit is properly taken, however, our results show that the bulk modulus, $B$, the shear modulus, $G$, and the ratio of the two, $G/B$, all become nonzero simultaneously at the jamming transition, consistent with earlier claims~\cite{OHern:2003vq}.

The location of the jamming transition depends on both system size~\cite{OHern:2003vq,Vagberg:2011fe} and protocol~\cite{Chaudhuri:2010jg}.  Thus, the packing fraction at the transition fluctuates from state to state.  Several studies have focused on finite-size effects associated with this distribution of packing fractions at the onset of jamming~\cite{OHern:2003vq,Vagberg:2011fe, Chaudhuri:2010jg,Liu:2014gu}.  In contrast, we concentrate on finite-size scaling in bulk quantities \emph{above} the transition, and bypass the effects of the distribution of jamming onsets by looking at behavior as a function of pressure, or equivalently, $\phi-\phi_c$, where $\phi_c$ is the packing fraction at the jamming onset for a given state.  

In Section~\ref{sec:jamming_def}, we introduce the three ensembles based on the different jamming criteria and review the constraint counting arguments for each one~\cite{Goodrich:2012ck,DagoisBohy:2012dh}.  We introduce the $\tfrac 18 d(d+1)(d^2+d+2)$ independent elastic constants in $d$ dimensions, and use them to find the conditions required for mechanical stability.  We then recast them in terms of combinations that either approach the bulk and shear moduli or vanish in the thermodynamic limit.  Section~\ref{sec:results} contains the numerical results for the excess contact number and the elastic constant combinations versus pressure and system size.  We also present results for statistical fluctuations of the excess contact number, bulk modulus, and $G_{DC}$.

\section{Jamming, ensembles and constraint counting in finite systems\label{sec:jamming_def}}

\begin{table}[h!]
\begin{tabular}{| c | l |}
\hline
Symbol 					& Meaning \\ \hline
$d$ 						& dimension \\
$N$ 						& total number of particles \\ 
$N_0$ 					& number of nonrattling particles \\
$\Nbndry$				& number of relevant boundary variables\\
$N_c$ 					& number of contacts \\
$\Nciso$					& isostatic number of contacts ($dN_0-d$) \\
$N_\text{c,min}$			& min number of contacts ($\Nciso + \Nbndry$) \\
$Z$ 						& contact number ($2 N_c/N_0$)\\
$\Ziso$ 					& isostatic contact number ($2d - 2d/N_0$)\\
$\Zmin$					& min contact number ($\Ziso +\frac 2{N_0} \Nbndry$) \\

\hline
$r_\alpha$ 				& position variable \\
$u_\alpha$ 				& displacement variable \\
$\Delta L_b$ 				& box shape variable \\
$V$ 						& volume \\
$\phi$ 					& packing fraction \\

\hline
$U$ 						& total energy \\
$\Delta U$ 				& change in energy \\
$p$ 						& pressure \\
$\epsilon_{ij}$				& strain tensor \\
$\sigma_{ij}$				& stress tensor \\
$s$						& residual shear stress\\
$H$						& enthalpy-like quantity ($U - \sigma_{ij}\epsilon_{ji}V$) \\
$\Delta H$				& change in $H$ \\
$\He^0_{\alpha,\beta}$		& regular Hessian matrix (dynamical matrix) \\
$\He_{\bar\alpha\bar\beta}$	& extended Hessian matrix \\
$q_{\bar \alpha}$			& union of $u_\alpha$ and $\epsilon_{ij}$ \\

\hline
$c_{ijkl}$					& elastic modulus tensor \\
$B$ 						& bulk modulus \\
$G$ 						& shear modulus of an isotropic system\\
$\theta$					& angle of a boundary deformation \\
$\hat \theta$				& generalized Euler angles in $d$ dimensions \\
$G(\hat\theta)$				& response to shear in direction $\hat \theta$ \\
$U(\hat\theta)$				& response to uniaxial compression in direction $\hat \theta$ \\
$D(\hat\theta)$				& dilatent response in direction $\hat \theta$ \\
$\Gdc$					& average response of a system to shear\\
$G_{AC}$					& stdev of response to shear \\
$U_{DC}$					& average response to uniaxial compression\\
$U_{AC}$					& stdev of response to uniaxial compression \\
$D_{DC}$					& average dilatent response (identically $0$) \\
$D_{AC}$					& stdev of dilatent response \\
$\sigma_X$				& stdev of $X_{DC}$ over ensemble ($X\in{G,U,D}$) \\
$B_0$					& zero pressure limit of $B$ \\
$G_{DC,0}$				& zero pressure limit of $G_{DC}$\\

\hline
\end{tabular}
\caption{\label{table:symbols}List of important symbols.}
\end{table}

\begin{table}
\begin{tabular}{| c | c | c | c |}
\hline
Index & Meaning & Range & {\it e.g.}  \\ \hline
$\alpha$, $\beta$ & particle position DOF & $[1,dN]$ & $r_\alpha$ \\
$\bar\alpha$, $\bar\beta$ & position and boundary DOF & $[1,dN+\Nbndry]$ & $q_{\bar\alpha}$ \\
$b$ & simulation box shape DOF & $[1, d(d+1)/2 -1]$ & $L_b$\\
$i$, $j$, $k$, $l$ & dimension & $[1,d]$ & $\epsilon_{ij}$\\
$n$ & mode number & $[1,dN]$ & $\lambda_n$ \\ \hline
\end{tabular}
\caption{\label{table:indices} List of indices and their meaning. Note that $d$ is the dimensionality and $N$ is the total number of particles.}
\end{table}

\subsection{Jamming Criteria and Ensembles\label{sec:JammingCriteria}}

We will consider athermal ($T=0$) packings of $N$ soft spheres that interact only when they overlap with a purely repulsive spherically symmetric potential in $d$ dimensions. For now, we will not be concerned with the specific form of the interaction potential and only require that it has a finite range that defines the particle diameter.
What does it mean for such a packing to be jammed? The answer to this is clear in the thermodynamic limit.
At sufficiently low packing fractions, $\phi$, there is room for the spheres to avoid each other so that none of them overlap, and the number of load-bearing contacts vanishes. The potential-energy landscape is locally flat and the pressure and elastic moduli, which are respectively related to the first and second derivatives of the energy, are zero; 
in no way should the system be considered a solid.
At high $\phi$, however, there is no longer room for the particles to avoid each other and they are forced to overlap, and 
the system possesses enough contacts for rigidity.
It no longer sits at zero energy and develops a non-zero stress tensor
with positive pressure. Moreover, the shear modulus $G$ and bulk modulus $B$ are positive. Such a system possesses all the characteristics of a solid and is therefore jammed.

When we are not in the large system limit, the onset of rigidity is more complex. In this section, we will discuss the behavior of three quantities -- the average contact number, the pressure and the elastic constants  -- in finite systems at the jamming transition.

{\em I: Connectivity ---} It has long been known that there is a connection between the jamming transition and the contact number $Z$ ({\it i.e.}, the average number of load-bearing contacts per non-rattling particle), which is given by $Z\equiv 2\Nc/N_0$, where $\Nc$ is the total number of contacts and $N_0$ is the number of particles that are not rattlers~\cite{BOLTON:1990uy,Durian:1995eo,Alexander:1998vc,Moukarzel:1998vn}.  $Z=0$ below the jamming transition because there are no overlapping particles. (Note, it is possible for two particles to \emph{just} touch, but such a contact cannot bear any load.) At the transition, $Z$ jumps to a finite value and increases further as the system is compressed. This finite jump has been understood from the Maxwell criterion, which is a mean-field argument stating that a rigid network of central-force springs must have an average contact number of at least $\Ziso$. 
When a system is isostatic ($Z=\Ziso$), 
the number of contacts just balances the number of degrees of freedom.

However, as pointed out in Ref.~\cite{Goodrich:2012ck}, the use of constraint counting and isostaticity as a measure of jamming has some serious drawbacks. For example, packings of ellipsoids jam well below isostaticity~\cite{Donev:2007go,Zeravcic:2009wo,Mailman:2009ct}.
Also, as contacts in frictional packings are able to constrain multiple degrees of freedom, the contact number at jamming depends sensitively on the strength of the frictional part of the interactions and lies below $2d$~\cite{Shundyak:2007ga,Somfai:2007ge,Henkes:2010uu,Henkes:2010kv,Papanikolaou:2013fa}.
%Also, some contacts in frictional packings are able to constrain multiple degrees of freedom, and the contact number at jamming depends sensitively on the strength of the frictional part of the interactions~\cite{Shundyak:2007ga,Somfai:2007ge,Henkes:2010uu,Henkes:2010kv,Papanikolaou:2013fa}. 
Furthermore, the Maxwell criterion assumes that as a system approaches isostaticity, none of the contacts are redundant (in a manner that can be defined precisely for certain networks). Although we will show below that this assumption is often correct, it is not a generic feature of sphere packings.

For example, consider a 50/50 mixture of large and small particles in two dimensions just above the jamming transition. Such bidisperse packings are quite common in the study of jamming because a monodisperse mixture leads to local crystallization. Even for bidisperse mixtures, however, there is a non-negligible probability that a particle is surrounded by 6 particles of exactly the same size. It is easy to see that these 7 particles have a redundant contact even at the transition, but this extra contact does not contribute to the global stability of the rest of the packing. Therefore, the contact number at the transition will be slightly greater than the isostatic value~\footnote{We find the difference to be small, of order $10^{-3}$.}. A corollary of this is that a packing might have $Z>\Ziso$ and still be unjammed. 
(As discussed in Appendix~\ref{sec:numerical_procedures}, our numerical calculations use a polydisperse distribution of particle sizes in two dimensions to avoid this issue.) Therefore, we see that constraint counting is not a robust indicator of whether or not a system is jammed. 

{\em II: Positive Pressure ---}
For packings of purely repulsive particles, positive pressure is clearly a necessary condition for jamming. If a particle is trapped by its neighbors, then there must be a restoring force to counteract any small displacement. Such forces can only come from particle-particle interactions which, when integrated over the system, lead to non-zero pressure. If the pressure is zero, then there cannot be any particle-particle interactions and the system is not jammed, regardless of system size. Therefore, positive pressure is a necessary condition for jamming.

{\em III: Mechanical Rigidity ---} A solid must resist global deformations such as compression and shear. We first consider the response to compression. As we saw above, particle-particle overlaps in a jammed system push outward and lead to non-zero pressure. Upon compression, these forces must increase to linear order, implying that the bulk modulus, $B$, is positive. 

\begin{figure}[htp]
\centering
\epsfig{file=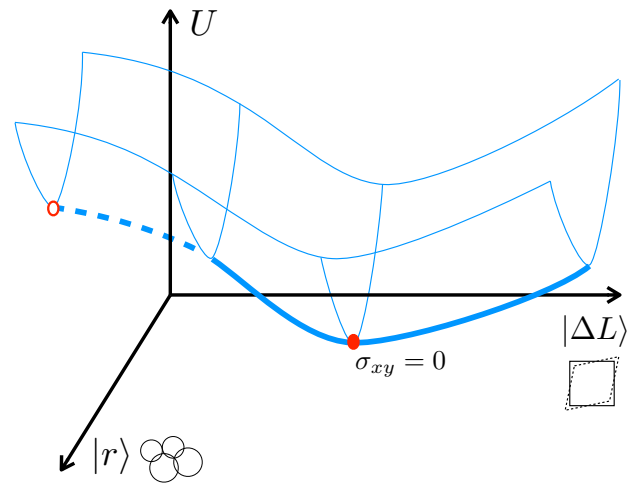,width=0.95\linewidth,clip}
\caption{\label{fig:energy_landscape}(color online).
Schematic energy landscape where $|r\rangle$ denotes the particle degrees of freedom and $|\Delta L \rangle$ all possible shear deformations of the box. 
For packings in the $\Ec$ ensemble, $|\Delta L\rangle$ is fixed and the system is $\Rc$ stable if it sits at a minimum of $U$ with respect to $|r\rangle$, {\it i.e.} the open circle. 
$\Ra$ stability is governed by the curvature of $U$ along the global shear degrees of freedom. 
Thus, $\Rc$ stable states can be $\Ra$ unstable if the curvature of $U$ is negative along any of the $|\Delta L\rangle$ directions (thick dashed curve). Such states can and do occur.
If the curvature of $U$ is positive along all global shear directions (thick solid curve), the packing is $\Ra$ stable. Such $\Ra$ stable packings can have finite shear stresses (non-zero gradient along global shear directions). Finally, packings that are at a local minimum of $U$ with respect to the $|r\rangle$ \emph{and} $|\Delta L\rangle$ directions (filled circle) have zero residual shear stress in addition to being $\Ra$ stable, and thus satisfy the $\Rap$ requirement.
}
\label{fig:SS_energy_landscape}
\end{figure}

The situation for shear deformations is more subtle, and various jamming criteria can be defined depending on the boundary conditions~\cite{DagoisBohy:2012dh}. 
Consider the potential energy landscape as a function of (1) the $dN$ particle positions $r_\alpha$, (2) the $d(d+1)/2 -1$ degrees of freedom $\Delta L_b$ associated with the shape of the box, and (3) the volume $V$. Common jamming algorithms fix the shape and size of the box and generate packings at a minimum of $U$ with respect to $\left| r \right> = \{r_\alpha\}$ (see Fig.~\ref{fig:SS_energy_landscape}).  In this case, no further constraints are necessary beyond those needed for the system to resist compression. 

The criterion that the system resist compression will be referred to as the $\Rc$, or ``Rigid to Compression,'' requirement, and the ensemble of systems that satisfy this requirement will be referred to as the $\Ec$ ensemble.  Experimental examples are when particles are placed in a rigid container or when the shape of the container is externally controlled.  Note that when the boundary is not allowed to deform, residual shear stresses and shear moduli correspond to the first and second derivatives, respectively, of $U$ along a strain direction without permitting the shape to equilibrate. As a result, such a system will generically have non-zero residual shear stresses.
%Such a system, once equilibrated, can have a residual shear stress. 
Likewise, as pointed out by Dagois-Bohy {\it et al.}~\cite{DagoisBohy:2012dh} and illustrated in Fig.~\ref{fig:energy_landscape}, systems that are $\Rc$ stable do not need to be stable to shear.
%As a result, both the residual stress and shear modulus are uncontrolled. 
% Fig.~\ref{fig:energy_landscape} illustrates that systems that are $\Rc$ stable do not need to be stable to shear, as pointed out by Dagois-Bohy {\it et al.}~\cite{DagoisBohy:2012dh}. 

The criterion that the system resists all global deformations, including shear and compression, will be referred to as the $\Ra$, or ``Rigid to All,'' requirement.  
As we will show below, an ensemble of systems that satisfy the $\Ra$ requirement can be obtained by filtering the $\Ec$ ensemble to keep only those systems that resist all global deformations. This ensemble will be referred to as the $\Ea$ ensemble. Previous work showed that the fraction of $\Ec$ packings that are $\Ra$ unstable becomes of order one for finite systems at sufficiently low pressure~\cite{DagoisBohy:2012dh}.

We can also consider the situation where the \emph{shape} of the container or simulation box is allowed to relax along with the particle positions~\cite{DagoisBohy:2012dh,Torquato:2010hb}. This introduces $d(d+1)/2 -1$ additional degrees of freedom, independent of system size, which are associated with the shape of the box. By expanding the dimensionality of the energy landscape, the system is able to relax to a lower energy minimum (see Fig.~\ref{fig:SS_energy_landscape}). Note that changing the shape of the simulation box can be interpreted as changing the metric tensor of the space in which the particles live~\cite{Torquato:2010hb}.

We have thus developed an algorithm for generating states that are not only $\Ra$ stable but also have zero residual shear stress~\cite{DagoisBohy:2012dh}.  In short, two-dimensional packings are generated by finding minima of $U$ with respect to both $\left| r \right>$ and the two shear degrees of freedom (labeled $\left| \Delta L \right> = \{\Delta L_b\}$ in Fig.~\ref{fig:SS_energy_landscape}). Because derivatives of $U$ with respect to shear degrees of freedom give shear stresses, the packings generated by this algorithm have a purely hydrostatic stress tensor.  Unlike algorithms that fix the shape of the simulation box, these packings are also guaranteed to have a positive shear modulus because the curvature of the energy landscape in the $\left| \Delta L \right>$ directions must be positive. We will refer to these combined criteria ($\Ra$ stable plus zero residual shear stress) as the $\Rap$ requirement. The ensemble of systems that satisfy the $\Rap$ requirement will be referred to as the $\Eap$ ensemble. 

\begin{figure}[tpb]
	\[\begin{tikzpicture}
		%make definitions
		\def\s{0.75}
		\def\xa{-2.1};	\def\ya{0};
		\def\xb{2.1};	\def\yb{0};
		
		\def\ria{2.5*\s}
		\def\rja{1.5*\s}
		\def\rka{0.6*\s}
		
		\def\rib{2.5*\s}
		\def\rjb{2.2*\s}
		\def\rkb{1.5*\s}
		
		%draw the circles
		\draw[pattern=north west lines, pattern color=blue!40, thick] (\xa,\ya) circle (\ria);
		\draw[pattern=north west lines, pattern color=blue!40,  thick] (\xb,\yb) circle (\rib);
		
		\draw[pattern=north east lines, pattern color=purple!30,  thick]  (\xa,\ya) circle (\rja);
		\draw[pattern=north east lines, pattern color=purple!30,  thick]  (\xb,\yb) circle (\rjb);
		
		\draw[pattern=grid, pattern color=green!!30,  thick] (\xa,\ya) circle (\rka);
		\draw[pattern=grid, pattern color=green!!30,  thick]  (\xb,\yb) circle (\rkb);
		
		\draw (\xa,\ya+2*\s) 	node {\Ec};
		\draw (\xa,\ya+1*\s) 	node {\Ea};
		\draw (\xa,\ya) 		node {\Eap};
		\draw[->, thick](\xb+2.*\s,\yb+2.2*\s) node[right]{$\!\!\Ec$} -- (\xb+1.64*\s,\yb+1.64*\s) ;
		\draw (\xb,\yb+1.8*\s) 		node {\Ea};
		\draw (\xb,\yb) 				node {\Eap};
		\draw (\xa,\ya+2.3)			node {small $pN^2$};
		\draw (\xb,\yb+2.3)			node {large $pN^2$};
	\end{tikzpicture}\]
	\caption{(color online).
		Schematic depiction of the hierarchical relation between the three ensembles $\Ec$, $\Ea$, and $\Eap$, for small $pN^2$ (left) and for
large $pN^2$ (right). Whereas for $pN^2 \rightarrow 0$, $\Ea$ becomes vanishingly small in comparison to $\Ec$~\cite{DagoisBohy:2012dh}, in the thermodynamic limit, $\Ec$ and $\Ea$ become virtually indistinguishable for all finite $p$. Moreover, for large systems the ratio of 
residual shear stress to pressure vanishes, so that the properties of $\Ea$ and $\Eap$ converge. 
	}
	\label{fig:diagram}
\end{figure}
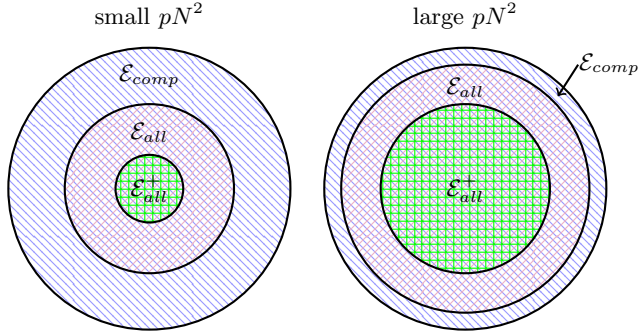

As illustrated in Fig.~\ref{fig:SS_energy_landscape}, these three jamming conditions have a simple interpretation in terms of the energy landscape. Furthermore, the ensembles have the hierarchical structure: $\Eap \subset \Ea \subset \Ec$ (see Fig.~\ref{fig:diagram}).

In the remainder of the paper we study three different ensembles of packings, the $\Ec$, $\Ea$ and $\Eap$ ensembles described above.
The standard $\Ec$ packings dominate the jamming literature; we study them in both two and three dimensions. We will refer to these as the ``2d \Ec" and ``3d \Ec" ensembles, respectively.  We will also study two dimensional packings that are $\Rap$ stable (stable to shear deformations in all directions \emph{and} have no residual shear stress), which make up the ``\Eap" ensemble. Finally, to compare these two ensembles, we consider the two-dimensional $\Ea$ ensemble, which is a ``filtered $\Ec$" ensemble where we include only the $\Ec$ configurations that happen to be $\Ra$ stable. Like the $\Eap$ states, $\Ea$ states have positive shear modulus; unlike the $\Eap$ states, $\Ea$ states have generically non-zero residual shear stress. 
The essential scenario is depicted in Fig.~\ref{fig:diagram}: whereas for small $p N^2$ the packings in these different ensembles are significantly different, for large $p N^2$ these differences become smaller and vanish when $p N^2 \rightarrow \infty$.
For further details and numerical procedures, see Appendix~\ref{sec:numerical_procedures}.

\begin{table}
\begin{tabular}{ | c | c | c | c |}
	\hline
	Ensemble & Criteria  & Preparation algorithm  & Dim \\ \hline
	$\Ec$ & $\Rc$  &standard jamming algorithm & $2d$, $3d$\\ \hline
	$\Ea$ & $\Ra$ & filtered $\Ec$ ensemble & $2d$ \\ \hline
	$\Eap$ & $\Rap$ & new shear-stabilized algorithm & $2d$ \\ \hline
\end{tabular}
\caption{\label{table:ensembles}List of ensembles, the jamming criteria they satisfy, the algorithm used, and the dimensionality in which we studied them: $\Eap \subset \Ea \subset \Ec$.  The distinction between these ensembles vanishes in the large-system limit.}
\end{table}

\subsection{Jamming criteria in terms of the extended Hessian\label{sec:precise_formulation}}
Here we show that the jamming criteria introduced in Sec.~\ref{sec:JammingCriteria} can be formulated in terms of an extended Hessian that includes the boundary degrees of freedom~\cite{DagoisBohy:2012dh,Tighe:2011fq}. 
By defining jamming in terms of global deformations, we avoid requiring that individual particles be constrained.  Assumptions about the existence of zero modes are also not required.  This formulation therefore avoids the ambiguities of previous definitions based on counting zero modes.
In practice, zero modes can be present in jammed systems, such as those associated with rattlers and the extended quartic modes in the zero pressure limit of jammed packings of ellipsoids~\cite{Donev:2007go,Zeravcic:2009wo,Mailman:2009ct} --- as long as they are decoupled from the boundary degrees of freedom, they do not prevent the packing from being jammed.

We will begin by considering the $\Ra$ requirement that the system be stable with respect to all possible boundary deformations, 
and then show how the less strict $\Rc$ requirement can be deduced in the same framework.
We start with the Taylor expansion of the potential energy $U$ about a reference state with energy $U^0$, volume $V^0$, and particles positions $r_\alpha^0$. We restrict our attention to reference states in which the sum of forces on each particle is zero. The goal will be to determine if the reference state is jammed.

To test the $\Ra$ requirement, we need to include the $\Nbndry = d(d+1)/2$ degrees of freedom associated with boundary deformations in the energy expansion. It will be convenient to represent these variables as a symmetric strain tensor, $\epsilon_{ij}$. By differentiating the energy with respect to $\epsilon_{ij}$, we get the stress tensor of the reference state:
\eq{	\sigma^0_{ij} = \frac{1}{V^0} \left( \frac{\partial U}{\partial \epsilon_{ij}}\right)_0 \, .	}
$\sigma^0_{ij}$ represents prestress in the system and the trace of $\sigma_{ij}^0$ is proportional to the pressure.

Now consider the set of $dN$ particle displacements $\{u_\alpha\}$ about the reference state, $u_\alpha \equiv r_\alpha - r^0_\alpha$. The net force on each particle is given by the derivative of the energy with respect to $u_\alpha$, but this must be identically zero to satisfy force balance. To treat the boundary deformations and particle displacements together, let $\{q_{\bar \alpha}\} = \{u_\alpha,\epsilon_{ij}\}$ be the combination of the $dN$ particle displacements and the $\Nbndry$ independent components of the strain tensor. The first order term in the energy expansion is $\left( \frac{\partial U}{\partial q_{\bar\alpha}} \right )_0 q_{\bar\alpha}$, but this reduces to $\sigma^0_{ij}\, \epsilon_{ji} \, V^0$ due to the presence of force balance.

If the boundary was held fixed, then the second order term in the expansion would be obtained from the Hessian matrix $\He_{\alpha\beta}^0$, which is given by 
\eq{	\He^0_{\alpha \beta} \equiv \left( \frac{\partial^2 U}{\partial u_\alpha \partial u_\beta} \right)_{\!0} \, ,	\label{eq:reduced_hessian}}
where the derivatives are evaluated at the reference state. $\He^0_{\alpha \beta}$ is also the well-studied dynamical matrix of a packing where every particle has unit mass; its eigenvectors give the normal modes of vibration. For perturbations that include the boundary, however, we instead need the ``extended Hessian" matrix $\hat K$ \cite{DagoisBohy:2012dh,Tighe:2011fq},
\eq{	\He_{\bar\alpha \bar\beta} \equiv \left( \frac{\partial^2U}{\partial q_{\bar\alpha}\partial q_{\bar\beta}} \right)_{\!0} \,.	
\label{eq:hessian}}
We refer to $\hat K$ as an extended Hessian due to the inclusion of the global degrees of freedom. 

To second order in $q$, the change in energy $\Delta U = U - U^0$ associated with a deformation is
\eq{
	\Delta U &\approx  \left( \frac{\partial U}{\partial q_{\bar\alpha}} \right )_0 q_{\bar\alpha}
		+ \frac{1}{2} \left( \frac{\partial^2 U}{\partial q_{\bar\alpha} \, \partial q_{\bar\beta}} \right )_0 q_{\bar\alpha} q_{\bar\beta} \nonumber \\
	    &\approx \sigma^0_{ij}\, \epsilon_{ji} \, V^0 + \frac{1}{2}  \hat K_{\bar\alpha \bar\beta} \, q_{\bar\alpha} q_{\bar\beta}  \,,
	\label{eq:expansion}
}
where the strain tensor $\epsilon_{ij}$ is determined from the last $\Nbndry$ components of $q_{\bar \alpha}$.  
The linear term represents work done against the pre-stress. Only the strain degrees of freedom contribute to the linear term; all other contributions sum to zero as a result of force balance in the reference state.

Two observations follow directly from the energy expansion of Eq.~(\ref{eq:expansion}). First, the presence of a linear term indicates that packings where force balance is satisfied on every particle do not generically sit at a minimum of their energy $U$ with respect to boundary deformations (Fig.~\ref{fig:energy_landscape}). Instead, gradients of the enthalpy-like quantity $H \equiv U -   \sigma^0_{ij} \, \epsilon_{ji} \, V^0$ vanish, $({\partial H}/{\partial q_{\bar\alpha}} )_0 = 0$: this requirement serves as a mechanical equilibrium condition.
Second, packings that are in {\em stable} $\Rap$ mechanical equilibrium under fixed confining stress must minimize $H$; this constrains the curvature of $\Delta H = H - H^0$, which is determined by the eigenvalues of the real and symmetric matrix $\hat K_{\bar \alpha \bar\beta}$. Packings that are only $\Ra$ stable do not minimize $H$ but still have the same constraints on the curvature of $\Delta H$. Defining $e_n$ and $\lambda_n$ to be the $n$th eigenvector and eigenvalue of $\He_{\bar\alpha\bar\beta}$, respectively,  we can write
\eq{
\Delta H = \frac{1}{2}  \hat K_{\bar\alpha \bar\beta} \, q_{\bar\alpha} q_{\bar\beta}  = \frac{1}{2} (q_{\bar\alpha} \, e_{n,\bar\alpha})^2 \lambda_n \,.
}
If $\lambda_n < 0$ for any mode, then the system is linearly unstable to perturbations along that mode. In this case, the system does not sit at a local energy minima and therefore is not jammed. 
In principle, zero modes ($\lambda_n=0$) are allowed, but if a zero mode has a non-zero projection onto any of the $\Nbndry$ boundary variables, then the system is unstable to that global deformation and again is not jammed. 
%If $\lambda_n = 0$, then 
%If such an unstable mode has any non-zero projection on any of the $\Nbndry$ boundary variables, then the system is unstable to that global deformation.

Therefore, for a system to be jammed according to the $\Ra$ requirement, it must satisfy
\eq{	\lambda_n \geq 0 \quad \forall n, 	\label{eq:jamming_def2}	}
and
\eq{	e_{n,\bar\alpha\p} &= 0 \quad \mbox{whenever $\lambda_n = 0$}, 	 \label{eq:jamming_def}	}
where $\bar\alpha\p$ runs only over the set of degrees of freedom associated with boundary deformations. Note that this definition automatically accounts for the presence of rattlers and the $d$ global translational zero modes.
%Thus, a system is jammed if $\He_{\bar\alpha \bar\beta}$ has no negative modes and if none of the zero modes couple to the boundary degrees of freedom.

For systems where the $\Rc$ requirement is the appropriate condition, jamming can be determined in much the same way. The only difference is in the relevant boundary variables and therefore the definition of the extended Hessian. Instead of considering all $d(d+1)/2$ boundary degrees of freedom, we only include isotropic compression/expansion. $\Nbndry = 1$ and the extended Hessian is thus a $dN+1$ by $dN+1$ matrix, but Eqs.~\eqref{eq:expansion}-\eqref{eq:jamming_def} follow identically.

For finite systems, the $\Ra$ requirement is significantly more strict than the $\Rc$ requirement. Packings made by standard jamming algorithms, which
are jammed according to the $\Rc$ requirement, can still have negative modes if shear deformations are included in the extended Hessian.  The fraction of states in the $\Ec$ ensemble that are also in the $\Ea$ ensemble is a function of $pN^2$ --- this fraction vanishes for small $pN^2$ but approaches 1 for large $pN^2$~\cite{DagoisBohy:2012dh}. This is depicted schematically in Fig.~\ref{fig:diagram}.

We stress that the definition in Eqs.~\eqref{eq:jamming_def2} and \eqref{eq:jamming_def} considers the eigenvalues and vectors of the extended Hessian defined in Eq.~(\ref{eq:hessian}). Although it is possible to calculate elastic moduli, and thus the stability, from the usual ``reduced'' Hessian of Eq.~\eqref{eq:reduced_hessian}~\cite{Maloney:2006dt}, the eigenvalues of the reduced Hessian are not sufficient to determine if a system is jammed. Indeed, a packing can be unstable to global deformations even when the reduced Hessian is positive semi-definite because positive (or zero) modes can become negative when they are allowed to couple to the boundary. 

\subsection{Jamming criteria in terms of elastic constants\label{sec:jamming_from_elastic_constants}}
The $\Rc$ and $\Ra$ requirements that a system be stable to boundary deformations are equivalent to placing restrictions on the elastic moduli.
For isotropic systems, where the elasticity is described by the bulk modulus, $B$, and the shear modulus, $G$, the connection between stability requirements and
elastic moduli is simple: the $\Rc$ requirement is satisfied when the bulk modulus is positive, while the $\Ra$ requirement is satisfied when both the bulk and shear moduli are positive.

However, finite-sized systems are not isotropic. As a result, individual packings with periodic boundary conditions should be treated as crystals with the lowest possible symmetry. In this section, we will discuss the elastic constants of such systems.

A global affine deformation is given to lowest order by a specific strain tensor $\epsilon_{ij}$, which transforms any vector $r_i$ according  to
\eq{ r_i \rightarrow r_i + \sum_j \epsilon_{ij} r_j. \label{strain_tensor_transformation}	}
Note that in $d$ dimensions, the strain tensor has $d(d+1)/2$ independent elements. Now, when a mechanically stable system is subject to an affine deformation, it usually does not remain in mechanical equilibrium. Instead, there is a secondary, non-affine response, which can be calculated within the harmonic approximation from the Hessian matrix discussed above. Details of this calculation are presented in Refs.~\cite{Ellenbroek:2006df,Ellenbroek:2009dp}.

The change in energy can be written as
\eq{	
	\frac {\Delta U}{V^0} =  \sigma^0_{ij}  \epsilon_{ji}  +  \frac 12 c_{ijkl}\epsilon_{ij}\epsilon_{kl},	
	\label{elastic_modulus_tensor_def}
 }
where $c_{ijkl}$ is the $d \times d \times d \times d$ elastic modulus tensor and $V^0$ is again the volume of the initial reference state. 
The symmetries of $\epsilon_{ij}$ imply:
\eq{	c_{ijkl} = c_{jikl} = c_{ijlk} = c_{klij}.	}
When no further symmetries are assumed, the number of independent elastic constants becomes $\tfrac 18 d(d+1)(d^2+d+2)$, which is 6 in 2 dimensions and 21 in 3 dimensions.

It is convenient to express Eq.~\eqref{elastic_modulus_tensor_def} as a matrix equation by writing the elastic modulus tensor as a symmetric $d(d+1)/2$ by $d(d+1)/2$ dimensional matrix $\tilde c$ and the strain tensor as a $d(d+1)/2$ dimensional vector $\tilde \epsilon$. In 2 dimensions, for example, these are
\eq{	
	\begin{array}{c c}
\tilde c = \left( \begin{array}{ccc}
\color{blue} c_{xxxx} & \color{PineGreen} c_{xxyy} & 2\color{Red} c_{xxxy} \\
. & \color{blue} c_{yyyy} & 2\color{Red} c_{yyxy} \\
. & . & 4\color{Green} c_{xyxy}
\end{array} \right),	&
\tilde \epsilon = \left( \begin{array}{c}
\epsilon_{xx} \\ \epsilon_{yy} \\ \epsilon_{xy}
\end{array} \right).
	\end{array}
	\label{C_colors_2d}
}
We can now rewrite Eq.~\eqref{elastic_modulus_tensor_def} as a matrix equation for the enthalpy-like quantity $\Delta H$:
\eq{	\frac{\Delta H}{V^0} = \frac 12 \tilde{\epsilon}^T \tilde{c}\,\tilde{\epsilon}. \label{DeltaU_matrixForm}}

We can now state the $\Rc$ and $\Ra$ requirements in terms of the anisotropic elastic moduli. The $\Rc$ requirement is that the system is stable against compression. This is measured by the bulk modulus, which can be written in terms of the elements of $c_{ijkl}$:
\eq{	B \equiv \frac 1{d^2} \sum_{k,l}c_{kkll}.	\label{B_def} }
The $\Rc$ requirement is satisfied if and only if $B>0$, which can be tested using Eqs.~\eqref{elastic_modulus_tensor_def} and \eqref{B_def}.

Unlike the bulk modulus, the shear modulus is not uniquely defined for anisotropic systems. Any traceless strain tensor constitutes pure shear, and to test the $\Ra$ requirement, we take a direct approach. The $\Ra$ requirement is satisfied if and only if $\Delta H > 0$ for all strain directions, {\it i.e.} for any $\epsilon_{ij}$. From Eq.~\eqref{DeltaU_matrixForm}, we see that this is the case if all the eigenvalues of $\tilde c$ are positive. Thus, the $\Ra$ requirement is satisfied if and only if $\tilde c$ is positive definite.

Note that the $\Rc$ and $\Ra$ requirements place different restrictions on the rank of $\tilde c$. For the $\Rc$ requirement, $\tilde c$ can have as few as one non-zero eigenvalue, while all $d(d+1)/2$ eigenvalues must be positive for the $\Ra$ requirement. This fact will be important in Sec.~\ref{sec:constraint_counting}.

\subsubsection{Useful elastic constant combinations\label{sec:usefulelasticconstants}}

Given the multitude of elastic constants, especially in higher dimensions, it is useful to divide them into 5 distinct types, based on their symmetry, as illustrated in Table~\ref{table:elastic_constant_types}. The most familiar are Types 1 and 2, which correspond to uniaxial compression and pure shear, respectively. 
For anisotropic systems, each elastic constant is independent and (generically) nonzero. However, our systems are \emph{prepared} under isotropic conditions; there is no {\it a priori} difference between any two axes, as there can be for crystals. Since the reference axes are arbitrary, we can rotate our coordinate system so that the elastic constant $c_{xxxx}$, for example, \emph{becomes} $c_{yyyy}$ in the new reference frame. The groups outlined in Table~\ref{table:elastic_constant_types} are defined so that any elastic constant can be rotated into another of the same type. They are thus conceptually equivalent, although of course their actual values will differ.

We will now exploit the conceptual distinction between the various types of elastic constants to define three orientation dependent moduli. A general description of this process is given in Appendix~\ref{AppendixA}, but for brevity we simply quote the results here. Let $\hat \theta$ be the set of generalized Euler angles that represent rotations in $d$ dimensions. The three $\hat \theta$-dependent moduli are the generalized shear modulus $G(\hat \theta)$, the modulus of uniaxial compression $U(\hat \theta)$, and the dilatancy modulus $D(\hat \theta)$. 
One could also construct an orientation dependent moduli for the Type 5 constants, but these only exist in three dimensions and will not be discussed here.
The bulk modulus is independent of orientation and is given by Eq.~\eqref{B_def}.

\begin{table}
\begin{tabular}{ | c | c | c | c |}
	\hline
	Type & Definition ($i \neq j \neq k$) & \# of constants & Example(s) \\ \hline
	1 & $c_{iiii}$ & $d$ & $c_{xxxx}$ \\ \hline
	2 & $c_{ijij}$ & $d(d-1)/2$ & $c_{xyxy}$ \\ \hline
	3 & $c_{iijj}$ & $d(d-1)/2$ & $c_{xxyy}$ \\ \hline
	4 & $c_{iiij}$, $c_{iijk}$ & $d^2(d-1)/2$ & $c_{xxxy}$, $c_{yyxz}$ \\ \hline
	5 & $c_{ijik}$ &$d(d-2)(d^2-1)/8$ &  $c_{xyxz}$ \\ \hline
\end{tabular}
\caption{\label{table:elastic_constant_types}Classification of elastic constants.}
\end{table}

As an example, consider the generalized shear modulus $G(\theta)$ in two dimensions. The set of symmetric, traceless strain tensors can be parameterized by the \emph{shear angle} $\theta$:
\eq{	\epsilon(\theta) = \frac \gamma 2 \left( \begin{array}{cc} \sin(2\theta) & \cos(2\theta) \\ \cos(2\theta) & -\sin(2\theta) \end{array} \right),	\nonumber}
where $\gamma \ll 1$ is the magnitude of the strain. When $\theta=0$, the response is given by $c_{xyxy}$, but when $\theta=\pi/4$, the response is $\tfrac 14(c_{xxxx}+c_{yyyy}-2c_{xxyy})$. For arbitrary angles, the response $G(\theta)$ is a sinusoidal function of $\theta$~\cite{DagoisBohy:2012dh} (see Appendix~\ref{AppendixA}).

\begin{figure}[htp]
	\centering
	\epsfig{file=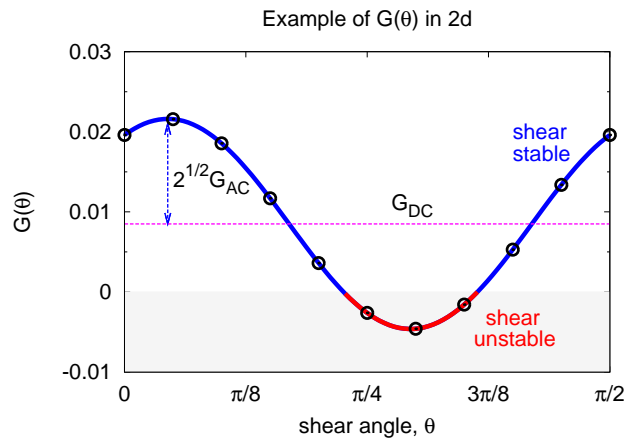,height=1\linewidth,angle=-90}
	\caption{\label{fig:G_theta}(color online).
	(a) Example of the sinusoidal function $G(\theta)$ for a two-dimensional system of $N=256$ particles at a pressure of $p=10^{-3}$. The black open circles show direct numerical calculations using various $\epsilon(\theta)$, while the solid line shows the prediction using the elastic modulus tensor at $\theta=0$ (see Appendix~\ref{AppendixA}). The system is stable to shear when $G(\theta)>0$ and unstable when $G(\theta)\leq 0$. The data agree with the prediction. The horizontal dashed line shows the average $\Gdc$; $G_{AC}$ is obtained from the amplitude of the sinusoidal curve. 
	}
\end{figure}

An example of $G(\theta)$ for a two-dimensional $\Ec$ packing is shown in Fig.~\ref{fig:G_theta}. Notice that there is a range of angles for which $G(\theta)<0$, implying that the system is unstable to that set of shear deformations. By construction, this does not occur for systems in the $\Ea$ and $\Eap$ ensembles. We define the angle-averaged shear modulus $\Gdc$ to be (see Fig.~\ref{fig:G_theta})
\eq{	G_{DC} &\equiv \frac{1}{\pi} \int_0^{\pi} G(\theta) d\theta.	}
We can also define $G_{AC}$ to characterize the variation of $G(\theta)$ about this average:
\eq{	G_{AC}^2 &\equiv \frac{1}{\pi} \int_0^{\pi} \left( G(\theta) - G_{DC} \right)^2 d\theta. \label{eq:G_AC_def}}

Note that for an isotropic system, $\Gdc = G$ ({\it i.e.}, the usual shear modulus) and $G_{AC}=0$. In three dimensions, the generalized shear modulus is no longer a simple sinusoidal function and instead depends on the three Euler angles. Nevertheless, we can still define $\Gdc$ and $G_{AC}$ to be the mean and standard deviation of the response to shear. This is discussed in detail in Appendix~\ref{AppendixA}.

In a similar manner, $U(\hat \theta)$ measures the response to uniaxial compression along an axis determined by $\hat \theta$. The full expression for $U(\hat \theta)$ is more complicated than $G(\hat \theta)$, but we can still define $U_{DC}$ and $U_{AC}^2$ to be the average and variance of $U(\hat \theta)$, respectively. However, since $U_{DC}$ can be expressed in terms of the bulk modulus and average shear modulus,
\eq{	U_{DC} = B + G_{DC}, \label{eq:Udc_from_B_and_G}}
it is redundant and will not be considered further. Finally, the Type 4 dilatancy constants can be generalized to $D(\hat \theta)$, and the average and variance defined as $D_{DC}$ and $D_{AC}^2$. One important result is that $D_{DC} = 0$ for any individual system (see Appendix~\ref{AppendixA}) and therefore will not be discussed further. 

In summary, we will consider the elastic constant combinations $B$, $G_{DC}$, $G_{AC}$, $U_{AC}$, and $D_{AC}$.  Expressions for these quantities in terms of the original elastic constants, $c_{ijkl}$, are provided in Appendix~\ref{AppendixA}.  Note that of these five quantities, $B$ and $G_{DC}$ reduce to the bulk and shear modulus, respectively, in the thermodynamic limit, which is isotropic.  As expected, we will see in Sec.~\ref{mvh2} that the remaining combinations,  $G_{AC}$, $U_{AC}$, and $D_{AC}$, vanish in the thermodynamic limit.

\subsection{Constraint Counting and Isostaticity\label{sec:constraint_counting}}
Earlier, we indicated that the contact number $Z$ is not an ideal metric for determining whether a system is jammed. However, the value of $Z$ at the jamming transition is of considerable importance. In this subsection, we review arguments from Ref.~\cite{Goodrich:2012ck} that derive the exact value of $Z$ at the jamming transition for packings of frictionless spheres in finite-sized systems in the $\Ec$ ensemble. In doing so, we also generalize the arguments to include the $\Ea$ and $\Eap$ ensembles and find that the contact number at the transition for these ensembles is slightly different~\cite{DagoisBohy:2012dh,Goodrich:2012ck}. This difference in contact number is easily understood from the additional degrees of freedom associated with boundary deformations that need to be constrained in the $\Ea$ and $\Eap$ ensembles. Furthermore, we will see in Sec.~\ref{sec:finite_size_scaling} that once this slight difference is taken into account, the increase in contact number with pressure is identical for the various ensembles. % (see Figs.~\ref{fig:finite_size_scaling} and \ref{fig:corrections_to_scaling}).

As discussed above, a system is isostatic when the number of constraints equals the number of degrees of freedom. Such a statement hides all subtleties in the definition of the relevant constraints and degrees of freedom.
For example, for a system with periodic boundary conditions in $d$ dimensions, particle-particle contacts cannot constrain global translational motion. Therefore, the isostatic number of contacts is
\eq{	\Nciso \equiv dN_0 - d,	}
where $N_0$ is the number of particles in the system after the rattlers have been ignored.  The isostatic contact number is therefore $\Ziso \equiv 2d - 2d/N_0$, which approaches $2d$ in the thermodynamic limit.

We now revisit the relationship between isostaticity and the jamming transition for packings of frictionless spheres. Suppose that $\Nzm$ of the total $dN+\Nbndry$ vibrational modes of the extended Hessian are zero modes, meaning they have zero eigenvalue. As before, $\Nbndry$ depends on the boundary conditions: $\Nbndry = 1$ in the $\Ec$ ensemble and $\Nbndry = d(d+1)/2$ in the $\Ea$ and $\Eap$ ensembles. A particle-particle contact has the potential  to constrain at most one degree of freedom, and every unconstrained degree of freedom results in a zero mode. Therefore, the number of contacts must satisfy
\eq{\Nc \geq dN + \Nbndry - \Nzm. \label{Nc_inequality} }
 Eq.~\eqref{Nc_inequality} is an inequality because some contacts might be redundant, meaning they could be removed without introducing a zero mode. Such redundancies correspond to states of self stress, and Eq.~\eqref{Nc_inequality} can be written as $\Nc = dN + \Nbndry - \Nzm + S$, where $S$ is the number of states of self stress~\cite{StenullLubensky}.

The $d$ global translations, as well as every rattler, each lead to $d$ trivial zero modes.
We will now use the numerical result that the only zero modes observed in jammed sphere packings are those associated with global translation and rattlers~\cite{Goodrich:2012ck}.  
Thus, the total number of zero modes in a jammed system is $\Nzm = d + d(N-N_0)$, and $\Nc$ and $Z$ must satisfy
\eqs{	
	\Nc & \geq N_\text{c,min} \equiv  \Nciso + \Nbndry, \\
	Z &\ge \Zmin \equiv \Ziso + \frac 2{N_0} \Nbndry.
	\label{Zmin_prediction}
}

If a system is exactly isostatic, then it has enough contacts to constrain the position of every particle, but it does \emph{not} have enough contacts to constrain the global degrees of freedom, and thus cannot be jammed. Since the $\Rc$ and $\Ra$ requirements do not explicitly forbid nontrivial zero modes, it is possible for the global variables to become constrained \emph{before} all the positional degrees of freedom. While this indeed occurs for ellipsoid packings, the fact that this is never observed for sphere packings implies that zero modes associated with translations of the spheres are extended and inevitably interact with the boundary.

\begin{figure}[htp]
	\centering
	\epsfig{file=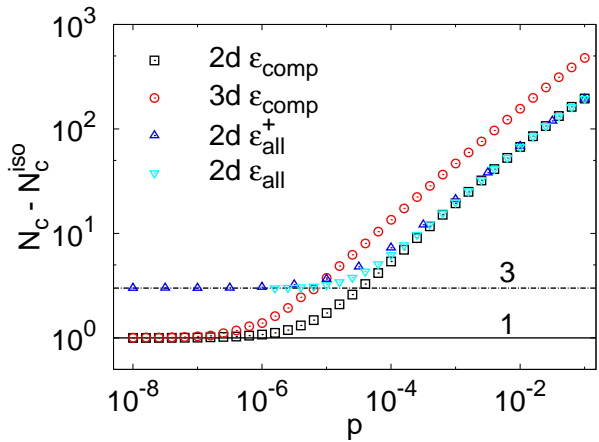,width=0.7\linewidth,angle=-90}
	\caption{(color online).
	The total number of contacts $\Nc$ above the isostatic number $\Nc^\text{iso}$ as a function of pressure for systems of $N=256$ particles. The solid horizontal line is at $\Nc-\Nciso=1$ and the dashed horizontal line is at $\Nc-\Nciso=3$.}
	\label{FigureNcmNciso}
\end{figure}

While Eq.~\eqref{Zmin_prediction} states that a system can only be jammed if $\Nc \geq \Nciso + \Nbndry$, this is clearly not a sufficient condition for jamming because some of the contacts could be redundant and not contribute to the overall rigidity of the system. However, we find numerically that Eq.~\eqref{Zmin_prediction} is indeed an equality as the transition is approached (provided the system is sufficiently disordered, recall the discussion in Sec.~\ref{sec:JammingCriteria}I regarding bidisperse packings in two dimensions). This is demonstrated in Fig.~\ref{FigureNcmNciso}, which shows that the number of contacts above isostaticity, in the limit of zero pressure, approaches $\Nc - \Nciso \rightarrow 1$ for the $\Ec$ ensembles (where $\Nbndry = 1$) and $\Nc - \Nciso \rightarrow 3$ for the $\Ea$ and $\Eap$ ensembles (where $\Nbndry = 3$). Importantly, we do not find \emph{any} systems that are jammed ({\it i.e.} satisfy Eqs.~\eqref{eq:jamming_def2} and~\eqref{eq:jamming_def}) but do not satisfy Eq.~\eqref{Zmin_prediction}.

Finally, note that the $\Nbndry$ additional contacts required for jamming can also be understood in terms of the normal reduced hessian and the matrix $\tilde c$ discussed in Sec.~\ref{sec:jamming_from_elastic_constants}. $\Nciso$ contacts are needed to remove any nontrivial zero modes from the reduced hessian. However, the $\Rc$ and $\Ra$ requirements necessitate $\Nbndry$ positive eigenvalues of $\tilde c$, leading to the additional $\Nbndry$ contacts in Eq.~\eqref{Zmin_prediction}.

\section{Numerical results\label{sec:results}}

\begin{figure*}[ht!pb]
	\centering
	\epsfig{file=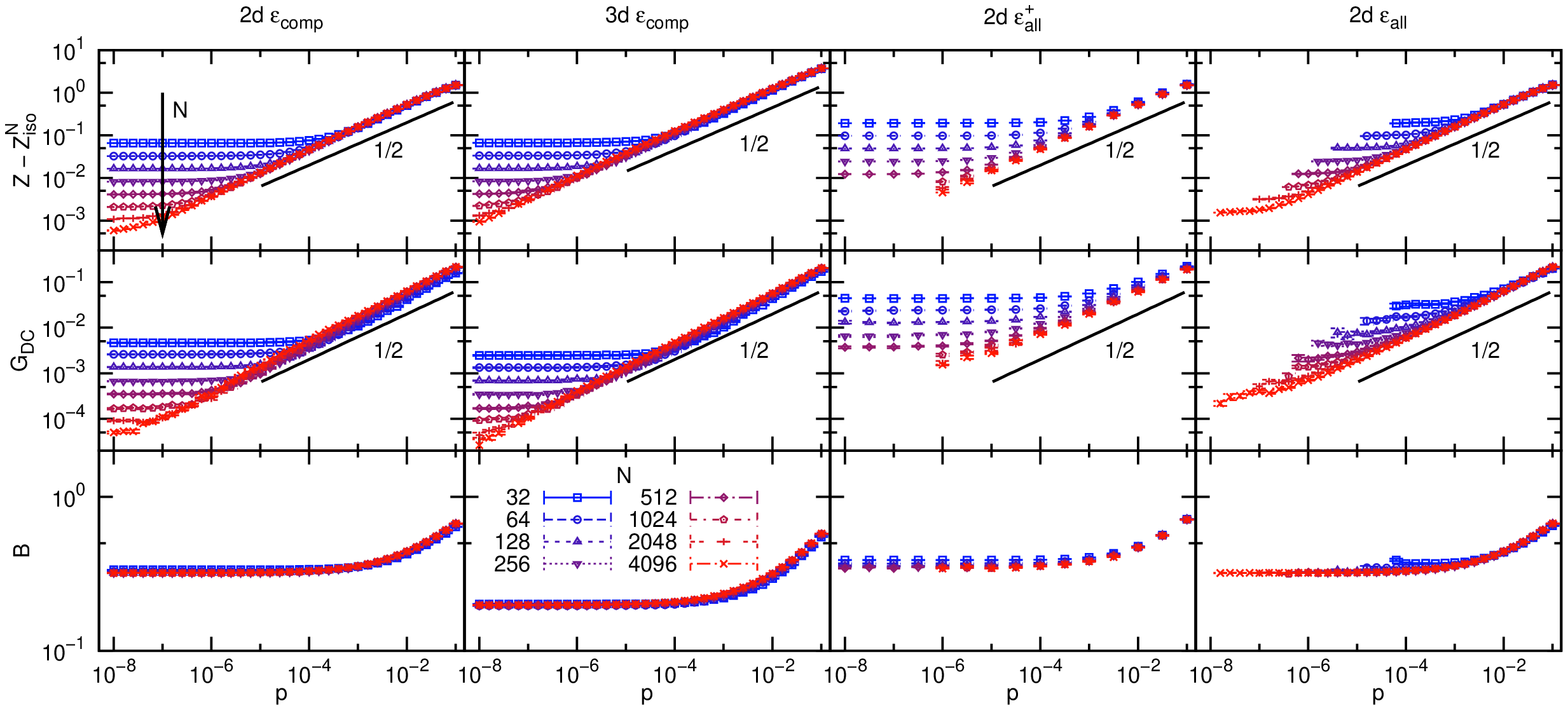,height=1\linewidth,angle=-90}
	\caption{(color online).
	The excess contact number ($Z - \Ziso$, top row), shear modulus ($G_{DC}$, middle row) and bulk modulus ($B$, bottom row) as a function of pressure for system sizes ranging from $N=32$ (blue squares) to $N=4096$ (red crosses). 
	The four columns show, from left to right, the $2d$ $\Ec$ ensemble, the $3d$ $\Ec$ ensemble, the $2d$ $\Eap$ ensemble, and the $2d$ $\Ea$ ensemble. The solid black lines all have slope $1/2$.}
	\label{FigureRawData}
\end{figure*}

In this section we examine the finite-size scaling behavior of the contact number and the elastic constants as a function of system size, $N$, and proximity to the jamming transition, which we quantify by the pressure, $p$, which vanishes at the transition. We will focus on soft-sphere potentials that have harmonic interactions (see Appendix~\ref{sec:numerical_procedures} for details), but extending our results to other soft-sphere potentials is straightforward~\cite{Liu:2010jx}. 

In Sec.~\ref{sec:finite_size_scaling} we present results for the excess contact number, $Z-\Ziso$, as well as for the two elastic constant combinations, $B$ and $G_{DC}$, that approach the bulk and shear moduli, respectively, in the thermodynamic limit (see Sec.~\ref{sec:usefulelasticconstants}).  Section~\ref{mvh2} contains the finite-size scaling results for the three ``$AC$" elastic constant combinations that vanish in the thermodynamic limit (again defined in Sec.~\ref{sec:usefulelasticconstants}).  Finally, Sec.~\ref{sec:statistical_fluctuations} examines the standard deviation of the distributions of the nonvanishing quantities, $Z-\Ziso$, $B$ and $G_{DC}$, namely $\sigma_Z$, $\sigma_B$ and $\sigma_{\Gdc}$.  These standard deviations must also vanish in the thermodynamic limit relative to the mean.  We note that when a single measurement of the response to shear, for example, is performed on a finite packing, both the angular variation and statistical fluctuations play a role ---
in earlier work we have shown examples where the angular and statistical fluctuations are taken together~\cite{DagoisBohy:2012dh}.

The results presented below can be summarized as follows.  First, we find subtle differences in $Z-\Ziso$, $B$ and $G_{DC}$ between the $\Ec$, $\Ea$ and $\Eap$ ensembles. These differences vanish as $pN^2 \rightarrow \infty$.  In addition, $G_{AC}$, $U_{AC}$, and $D_{AC}$ all vanish in the thermodynamic limit, as expected, and the fluctuations, $\sigma_Z$, $\sigma_B$ and $\sigma_{\Gdc}$, all vanish as $1/\sqrt{N}$ relative to the mean. All 6 quantities that vanish in the thermodynamic limit ($G_{AC}$, $U_{AC}$, and $D_{AC}$, $\sigma_Z$, $\sigma_B$ and $\sigma_{\Gdc}$) collapse with $pN^2$ in all 3 ensembles, with the exception of $U_{AC}$ and $D_{AC}$, which only collapse in the $\Eap$ ensemble, where there is no residual shear stress.  We will discuss these two exceptions further below.  In all, these results show that the thermodynamic limit is well defined for any $p$, although the number of particles needed to observe this limit diverges as the jamming transition is approached.  

Second, we find non-trivial finite-size corrections to the scaling of $Z-\Ziso$, $B$ and $G_{DC}$, in all three ensembles, as found for the $\Ec$ ensemble earlier~\cite{Goodrich:2012ck}. 
These corrections scale with the total system size, $N$, rather than the system length, $L$, in 2 and 3 dimensions, consistent with Ref.~\cite{Goodrich:2012ck}.  In addition, we find that the two-dimensional results can be better described when logarithmic corrections to scaling are included.  These results therefore reinforce the conclusion that  jamming is a phase transition with an upper critical dimension of two.

\subsection{Finite-Size Scaling: $Z-\Ziso$, $G_{DC}$ and $B$\label{sec:finite_size_scaling}}

In this section we probe the finite-size scaling of the ensemble-averaged values of the angle-independent quantities that do not vanish in the thermodynamic limit:
the contact number above isostaticity, $Z-\Ziso$, the shear modulus, $G_{DC}$, and the bulk modulus, $B$.  We study these for all three ensembles defined earlier.

\subsubsection{Finite-Size Plateau}

Figure~\ref{FigureRawData} shows the excess contact number, $Z- \Ziso$, average shear modulus,
$G_{DC}$, and bulk modulus, $B$, as a function of pressure for different system sizes and ensembles. At high pressures we measure the scaling relationship $Z-\Ziso \sim p^{1/2}$ that has previously been observed~\cite{Durian:1995eo,OHern:2003vq,Liu:2010jx} for harmonic interaction potentials. However, at low pressures the excess contact number plateaus to $2\Nbndry/N_0$. As expected, this correction to the excess contact number due to stabilizing the boundaries is a finite-size effect: as the system size increases, the onset pressure of this plateau decreases so that $Z-\Ziso \sim p^{1/2}$ is valid for all pressures in the thermodynamic limit. 

Similar to the excess contact number, the shear modulus has a high-pressure regime that conforms to the known scaling of $G_{DC} \sim p^{1/2}$ and a low-pressure plateau that scales as $1/N$. This plateau also vanishes in the thermodynamic limit and is a finite-size effect. 
The nearly constant behavior of the bulk modulus as a function of pressure is consistent with previous results and persists for large systems.

The fact that $\Gdc/B$ in the limit of zero pressure is proportional to $1/N$ and thus vanishes for large systems is a key feature of the jamming transition. In random spring networks, which are often used to model disordered solids, both the shear and bulk moduli vanish when the system approaches isostaticity such that the ratio of the two remains finite~\cite{Ellenbroek:2009to}. The only model system we are aware of that exhibits this jamming-like behavior in $\Gdc/B$ is the set of ``generic" rational approximates to the quasi-periodic Penrose tiling. In recent work~\cite{StenullLubensky}, Stenull and Lubensky show that such networks near isostaticity have constant bulk modulus (for sufficiently large $N$) and a shear modulus that vanishes with $1/N$. Their results are also consistent with our discussion in Sec.~\ref{sec:constraint_counting}.

\subsubsection{Finite-size scaling of excess contact number, bulk and shear moduli}

\paragraph{Contact Number:}
If jamming is a phase transition, then quantities like the excess contact number, $Z-\Ziso$, must be analytic for finite $N$. However, the bulk scaling of $Z-\Ziso \sim p^{1/2}$ that has been known for over a decade~\cite{Durian:1995eo,OHern:2003vq} is clearly not analytic at $p=0$. Thus, there must be finite-size rounding of this singular behavior if jamming is to be considered critical. For example, we already saw that finite-size effects in $Z-\Ziso$ emerge in the limit of zero pressure, resulting in a plateau that is proportional to $1/N$. Criticality also implies that such finite-size rounding should exhibit scaling collapse. Here, we will use the assumptions of finite-size scaling and analyticity at $p=0$, along with our understanding of the low-pressure plateau and the high-pressure scaling, to extract the scaling form and predict an additional finite-size effect that cannot be understood from constraint counting alone. This prediction is that for small $pN^2$, the increase in the contact number above its minimum is proportional to $pN$. We then numerically confirm this prediction as well as the initial assumption that finite-size scaling exits. These arguments were presented in an abbreviated form in Ref.~\cite{Goodrich:2012ck}, and are included with more detail here.

%If jamming is a phase transition, then finite-size effects in quantities like the excess contact number, $Z-\Ziso$, should exhibit scaling collapse. Furthermore, the scaling function obtained from finite-size scaling must be analytic for finite $N$ since true singularities can only occur in the thermodynamic limit. However, the bulk scaling of $Z-\Ziso \sim p^{1/2}$ that has been known for over a decade~\cite{Durian:1995eo,OHern:2003vq} is clearly not analytic at $p=0$. Thus, there must be finite-size rounding of this singular behavior if jamming is to be considered critical. We already saw that finite-size effects in $Z-\Ziso$ emerge in the limit of zero pressure, resulting in a plateau that is proportional to $1/N$. Here, we predict additional finite-size effects that cannot be understood from constraint counting alone. Instead, we will use our theoretical understanding of the low-pressure plateau and the high-pressure scaling to infer the scaling of more subtle finite-size effects that arise from the assumption that $Z-\Ziso$ is analytic for finite $N$.  These arguments were presented in an abbreviated form in Ref.~\cite{Goodrich:2012ck}, and are included with more detail here.

\begin{figure*}[h!t]
	\centering
	\epsfig{file=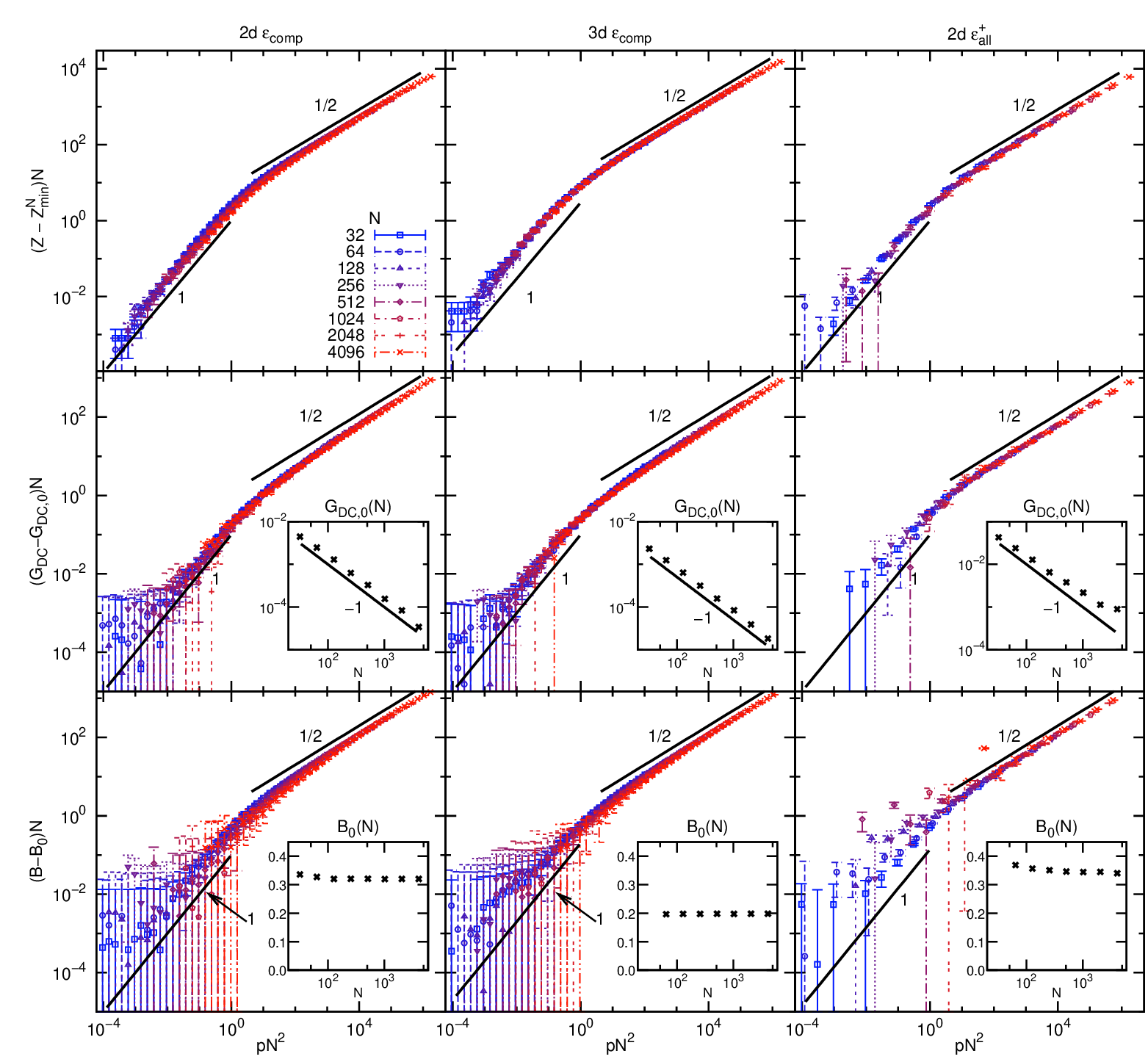,height=0.95\linewidth,angle=-90}
	\caption{\label{fig:finite_size_scaling}(color online).
	Finite-size scaling collapse of the excess contact number and average elastic moduli. Top row: $Z$ minus the theoretical minimum $\Zmin$ (see Eq.~\eqref{Zmin_prediction}). $\left(Z-\Zmin\right)N$ collapses as a function of $pN^2$ for the 2d $\Ec$ (left), 3d $\Ec$ (middle) and 2d $\Eap$ (right) ensembles. Note that at low $pN^2$, most of the $\Ec$ packings are $\Ra$ unstable and our filtered, 2d $\Ea$ ensemble does not have many states at low $pN^2$. This is why data is not shown for this ensemble. At large $pN^2$, $Z-\Zmin \sim Z-\Ziso \sim p^{1/2}$ (Eq.~\eqref{ZmZiso_p_scaling}), while $Z-\Zmin \sim pN$ at low $pN^2$ (Eq.~\eqref{ZmZmin_predicton}). The crossover between these scalings occurs when the total number of extra contacts is of order 10.
	Middle row: $\Gdc$ minus the measured $p\rightarrow 0$ plateau. $\left(\Gdc - G_{DC,0}\right)N$ collapses as a function of $pN^2$ and has the same crossover behavior as $\left(Z-\Zmin\right)N$. The insets show that $G_{DC,0}$ is proportional to $N^{-1}$.
	Bottom row: $B$ minus the measured $p\rightarrow 0$ plateau. Note that the plateau $B_0$ of the bulk modulus is much larger than for the shear modulus. Therefore, uncertainties in $B$ lead to the large error bars in $\left(B- B_0\right)N$ at low $pN^2$. The insets show that $B_0$ is roughly constant in $N$, as expected for particles with harmonic interactions. It is not clear from the data whether there is an additional $N^{-1}$ contribution to the plateau ({\it i.e.} $B_0(N) = B_0(\infty) + a N^{-1}$).
	The colors and symbols are the same as in Fig.~\ref{FigureRawData}.
	}
\end{figure*}

\begin{figure}[htp]
	\centering
	\epsfig{file=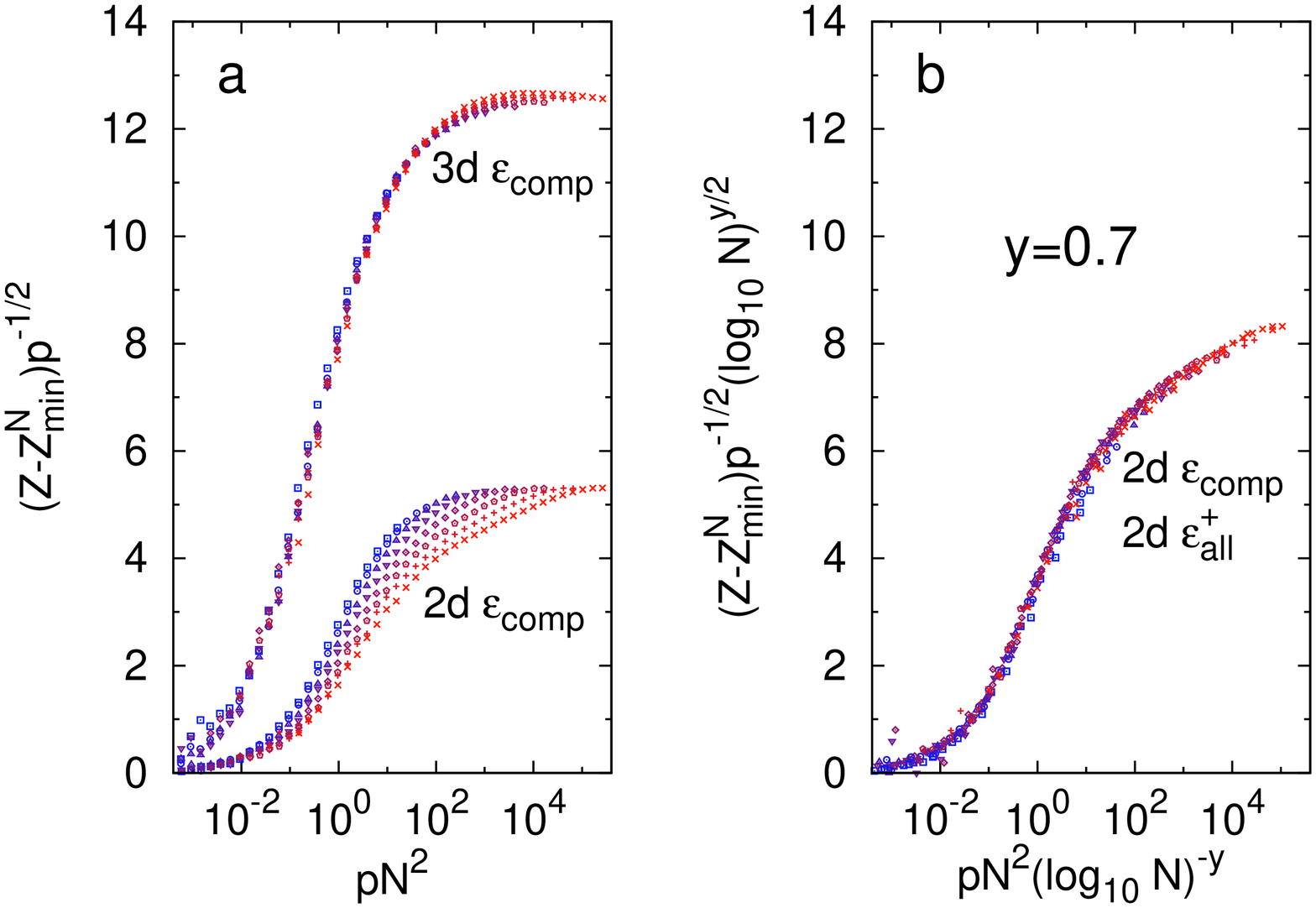,height=1\linewidth,angle=-90}
	\caption{\label{fig:corrections_to_scaling}(color online).
	a) $\left(Z-\Zmin\right)p^{-1/2}$ as a function of $pN^2$ on a linear scale for the two and three dimensional $\Ec$ ensembles. The $3d$ data shows good collapse but there is a system size dependence in $2d$ that was not as clear in Fig.~\ref{fig:finite_size_scaling}. The two dimensional $\Eap$ data (not shown) is indistinguishable from the 2d $\Ec$ data. b) $\left(Z-\Zmin\right)p^{-1/2}\left(\log_{10} N\right)^{y/2}$ as a function of $pN^2\left(\log_{10} N\right)^{-y}$, with $y=0.7$. The 2d $\Ec$ and 2d $\Eap$ ensembles are both shown and collapse perfectly onto each other. This shows that the system size dependence in a) can be accounted for by introducing a logarithmic correction of the form of Eq.~\eqref{scaling_form_log_corrections}. The colors and symbols are the same as in Fig.~\ref{FigureRawData}.}
\end{figure}

%The assertion that jamming is a phase transition leads to a falsifiable prediction for the form of the scaling function.  
First we summarize the three main ingredients of the argument. {\em(i)} The low pressure plateau in $Z-\Ziso$ derives from the extra contact(s) needed to satisfy the jamming criteria and is proportional to $1/N$. {\em(ii)} In the limit of large $N$ and at sufficiently large pressures, $Z-\Ziso$ exhibits power-law scaling with a known exponent of $1/2$:
\eq{	Z-\Ziso \sim p^{1/2}.	\label{ZmZiso_p_scaling} }
{\em(iii)} $Z$ is analytic in $p$ for finite $N$.

From the first two assertions, we see that if finite-size scaling is obeyed, it must be of the form
\eq{	Z-\Ziso = \frac 1N F\left( pN^2\right),	\label{scaling_form} }
where $F(x)$ is a scaling function that must satisfy, first, that
$F(x)\sim 1$ for small $x$, second, that $F(x) \sim x^{1/2}$ for large $x$, and third, that $F(x)$ is analytic in $x$ at $x=0$.

The third requirement regarding analyticity implies that
the expansion of the contact number for small $p$ takes the form
\eq{	\left(Z-\Ziso\right)N = c_0 + c_1 pN^{2} + ..., \label{ZmZiso_expansion}}
where $c_0 = 2\Nbndry$ gives the zero pressure plateau and $c_1$ is a constant.
Although the leading terms in the expansion clearly fail to describe the $Z-\Ziso \sim p^{1/2}$ scaling at large pressure, they should be valid at small pressure.
Our reasoning thus predicts that as the pressure vanishes, the contact number should approach its limiting value $\Zmin$ as
\eq{	Z-\Zmin \approx c_1 pN \qquad \text{for $p \ll 1$},	\label{ZmZmin_predicton} }
where the constant $c_1$ is independent of system size. Furthermore, there should be a crossover between this low-pressure regime and a high-pressure regime where $Z-\Zmin \sim Z-\Ziso \sim p^{1/2}$.

This is verified in the top row of Fig.~\ref{fig:finite_size_scaling}, which shows that $\left(Z-\Zmin\right)N$ does indeed collapse as a function of $pN^2$. The scaling with exponent $1/2$ at high $pN^2$ is consistent with Eq.~\eqref{ZmZiso_p_scaling}, while the slope of $1$ at low  $pN^2$ is consistent with Eq.~\eqref{ZmZmin_predicton}. Since $\left(Z-\Zmin\right)N$ is exactly twice the total number of contacts above the minimum ({\it i.e.}, $\Nc - N_\text{c,min}$), our data shows that the crossover to the low-pressure regime occurs when the total number of extra contacts in the system is of order 10, regardless of the system size. Importantly, the low-pressure scaling is not predicted from constraint counting arguments and data collapse in this region is not trivial. However, both follow immediately from the notion that jamming is a phase transition.

\paragraph{Shear Modulus:}
We now turn our attention to the average shear modulus $\Gdc$. We saw in Fig.~\ref{FigureRawData} that the behavior of $\Gdc$ is strikingly similar to that of $Z-\Ziso$. Specifically, the shear modulus deviates from the canonical $\Gdc \sim p^{1/2}$ scaling at low pressure and instead exhibits a plateau that decreases with system size. As we discussed above, this plateau is due to the $\Rc$ and $\Ra$ requirements that there are at least $\Nbndry$ constraints above the isostatic value.

Since $Z -\Ziso \sim N^{-1}$ in the zero-pressure limit, one would also expect the plateau in $\Gdc$ to be proportional to $N^{-1}$. Using the same reasoning as above, if finite-size scaling exists in the shear modulus it must be of the form $\Gdc N \sim  F(pN^2)$, where again $F(x)\sim 1$ for small $x$ and $F(x) \sim x^{1/2}$ for large $x$.
Also, the assertion that $\Gdc$ is analytic for finite $N$ implies that the low-pressure limit of the shear modulus is of the form
\eq{	\Gdc N = g_0 + g_1 pN^{2} + ...}
where $g_0$ and $g_1$ are constants.

The middle row of Fig.~\ref{fig:finite_size_scaling} confirms this scaling. For each ensemble and system size, we first calculated the plateau value $G_{DC,0}$ of $\Gdc$, and then plotted $\left(\Gdc-G_{DC,0}\right)N$ as a function of $pN^2$. The values of $G_{DC,0}$ are shown in the insets and are proportional to $N^{-1}$, confirming that $g_0$ is indeed constant.
$\Gdc$ increases from this plateau at low pressures with $pN$ before crossing over to the known $p^{1/2}$ scaling.

\paragraph{Bulk Modulus:} The same reasoning as above can also be applied to the scaling of the bulk modulus. As the bottom row of Fig.~\ref{fig:finite_size_scaling} shows, our data appear consistent with $(B-B_0)N$ scaling linearly with $pN^2$ close to the transition.  However, the error bars are very large as the plateau value for the bulk modulus is orders of magnitude larger than that of the shear modulus so the bulk modulus does not supply nearly as strong support for the existence of nontrivial scaling as the shear modulus and coordination number.

The finite-size effects presented in Figs.~\ref{FigureRawData} and \ref{fig:finite_size_scaling} clearly depend on the pressure, which is a useful measure of the distance to jamming for an individual system. 
A recent paper~\cite{Liu:2014gu}, however, claims to see finite-size scaling of the contact number and shear modulus with $(\phi-\phi_{c,\infty})L^{1/\nu}$, where $\nu \approx 0.8$, which is the same scaling that controls the mean of the distribution of critical packing fractions~\cite{OHern:2003vq,Vagberg:2011fe}.
To understand this, note that there are two different finite-size effects that come into play: 1) the corrections to $\phi_c$ that scale with $(\phi-\phi_{c,\infty})L^{1/\nu}$, and 2) the rounding shown in Figs.~\ref{FigureRawData} and \ref{fig:finite_size_scaling} that scale with $pN^2\sim pL^{2d}$. Since $1/\nu<2d$, one would expect the corrections to $\phi_c$ to influence the contact number and shear modulus over a broader range of $\phi$, leading to the observations of Ref.~\cite{Liu:2014gu}. However, the true behavior of these quantities as a function of $\phi$ is a convolution of the two finite-size effects. Thus, given their different scaling, finite-size collapse can not exist as a function of $\phi$. The appearance of scaling collapse observed in Ref.~\cite{Liu:2014gu} is because their data is not sufficiently sensitive at low pressures.

\subsubsection{Corrections to scaling in two dimensions}

We return now to the scaling for the contact number, and note the quality of the data collapse in three dimensions, which spans over 8 decades in $pN^2$ and over 5 decades in $\left(Z-\Zmin\right)N$ (see Fig.~\ref{fig:finite_size_scaling}). In both of the two-dimensional ensembles, however, there is a very slight systematic trend at intermediate $pN^2$. This can be seen more clearly by dividing $\left(Z-\Zmin\right)N$ by $p^{1/2}N$ and showing the data on a linear scale. Figure~\ref{fig:corrections_to_scaling}a shows that the collapse of the $3d$ data remains extremely good while there are clear deviations in the $2d$ data.

These deviations can be interpreted as corrections to scaling, which are often observed in critical phenomena at the upper critical dimension. 
One would expect potential corrections to scaling to be logarithmic and lead to scaling of the form
\eq{	Z-\Ziso = \frac 1N F\left( pN^2/\left(\log N\right)^y \right),	 \label{scaling_form_log_corrections} }
with some exponent $y$. Figure~\ref{fig:corrections_to_scaling}b shows both the 2d $\Ec$ data and the 2d $\Eap$ data scaled according to Eq.~\eqref{scaling_form_log_corrections}. We find that including a logarithmic correction with $y = 0.7 \pm 0.1$ leads to very nice data collapse in two dimensions.

The finite-size scaling that we observe depends on the total number of particles $N$ rather than the linear size of the system $L\sim N^{1/d}$. Such scaling is typically associated with first-order transitions and with second-order transitions above the upper critical dimension~\cite{BINDER:1985vl,Dillmann:1998ty}. Along with the corrections to scaling that we see in $d=2$, this is consistent with the notion that jamming is a mixed first/second order phase transition with an upper critical dimension of $d_\text{c} = 2$, in accord with previous results
~\cite{Liu:2010jx,Charbonneau:2012fl,Goodrich:2012ck,Wyart:2005vu}.

Unlike $\left(Z-\Ziso\right)N$, which approaches the same small pressure plateau in every individual system, the plateaus in $\Gdc$ vary from system to system. It is only when averaged over many systems that $G_{DC,0}$ has a clear $N^{-1}$ scaling. This explains why $(\Gdc - G_{DC,0})N$ is much more noisy at low $pN^2$ than $(Z-\Zmin)N$, which makes it impossible to see from our data whether or not there are corrections to scaling in $\Gdc$ in two dimensions.

\begin{figure*}[htp]
	\centering
	 \epsfig{file=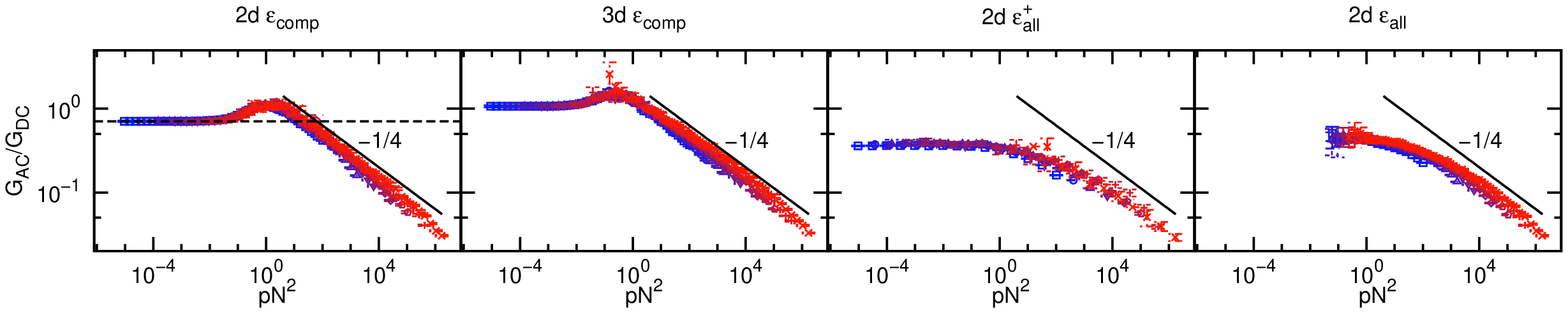,height=1\linewidth,angle=-90}\\
	\epsfig{file=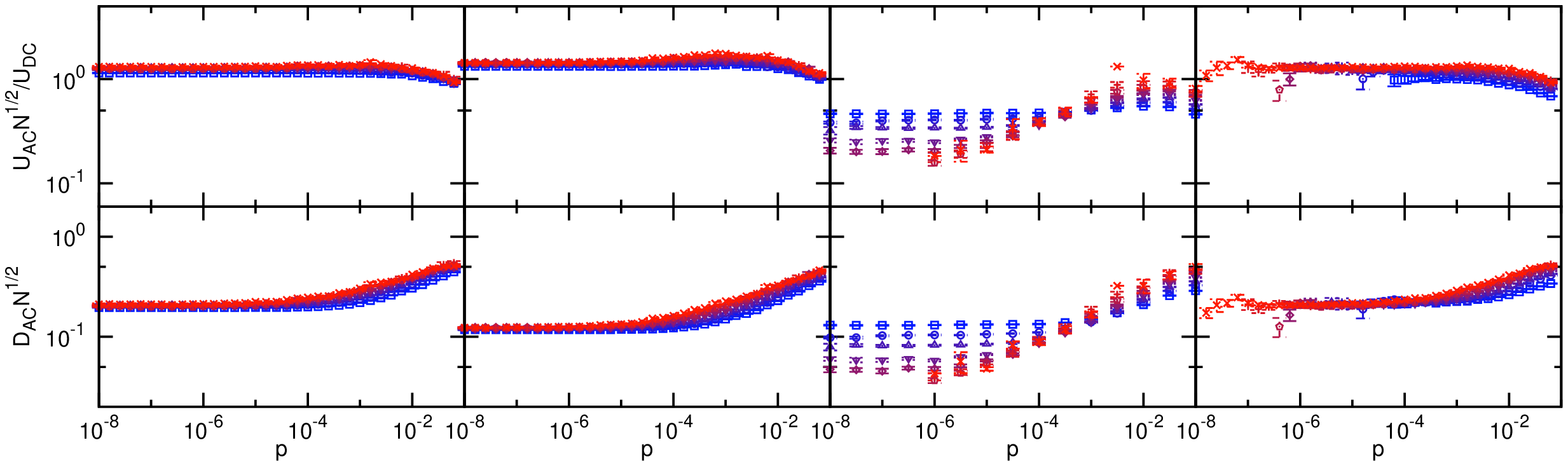,height=1\linewidth,angle=-90}
	\caption{\label{fig:anisotropic_AC_scaled}(color online).
	The average ``$AC$" quantities, which are defined in Appendix~\ref{AppendixA} and discussed in the text. The colors and symbols are the same as in Fig.~\ref{FigureRawData}.}
\end{figure*}

\begin{figure}[htp]
	\centering
	 \epsfig{file=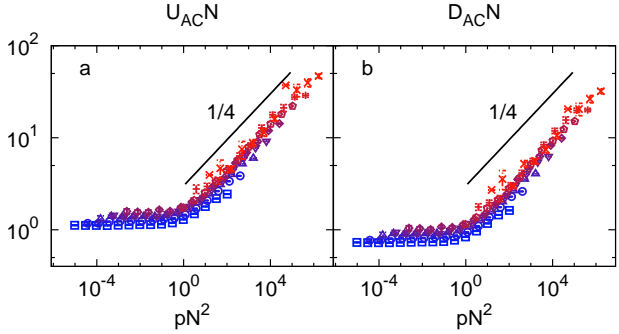,height=1\linewidth,angle=-90}
	\caption{\label{fig:anisotropic_AC_scaled_SS}(color online).
	Scaling collapse of $U_{AC}$ and $D_{AC}$ for the 2d $\Eap$ ensemble. The scaling of these two quantities is the only unexpected difference between the three ensembles that we have found. The colors and symbols are the same as in Fig.~\ref{FigureRawData}.}
\end{figure}

\begin{figure}[htp]
	\centering
	 \epsfig{file=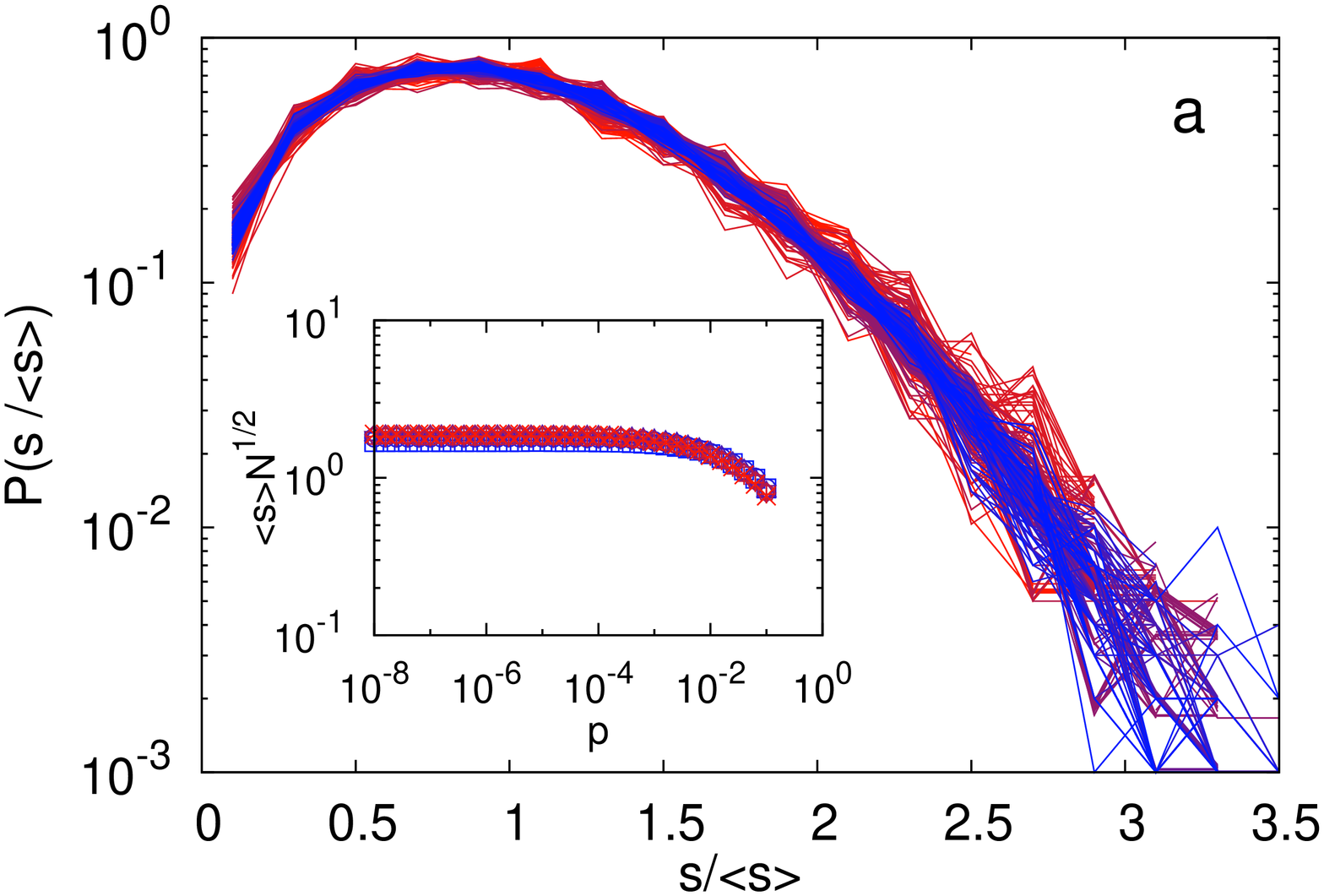,height=0.8\linewidth,angle=-90}
	 \epsfig{file=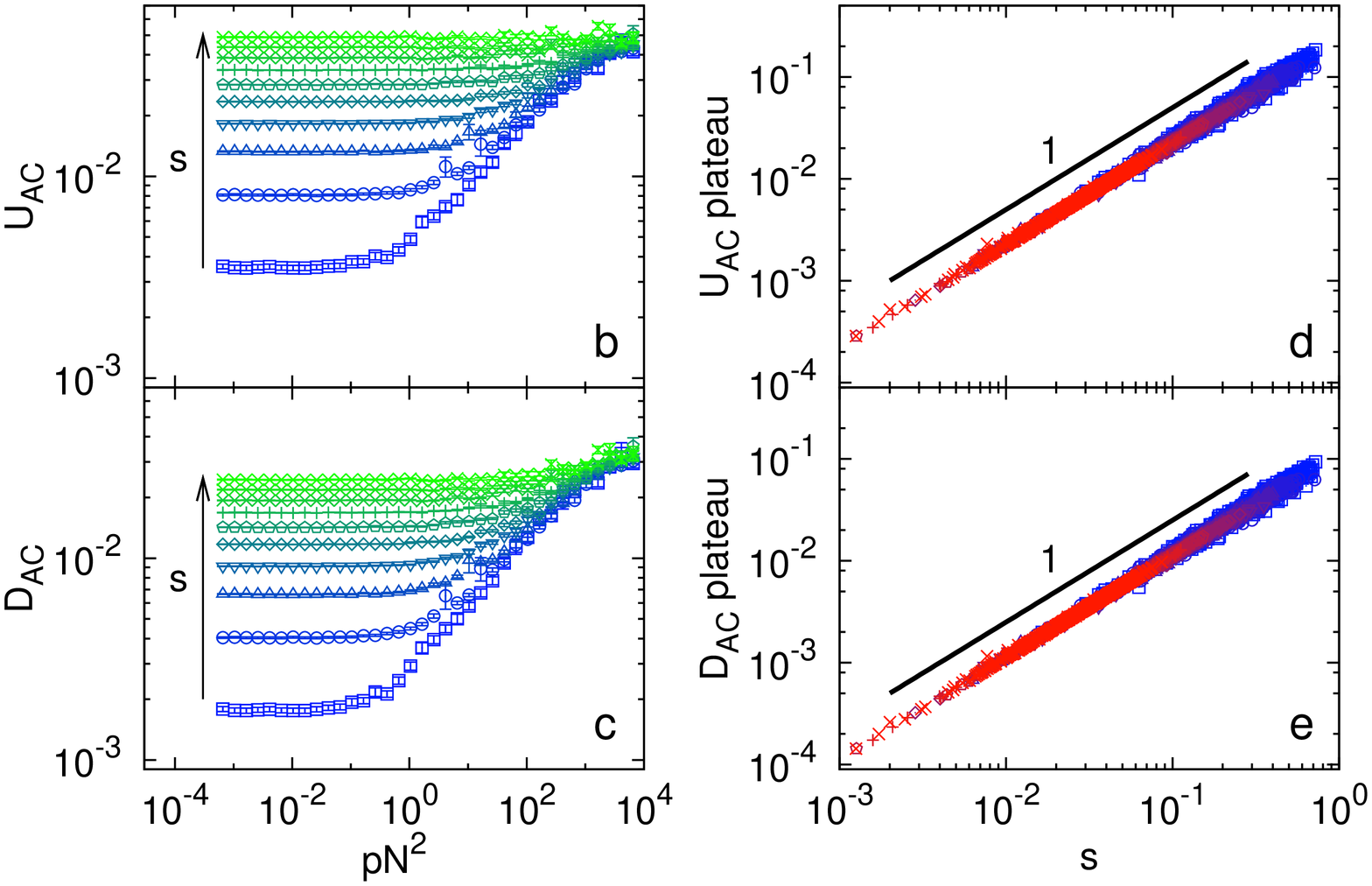,height=1\linewidth,angle=-90}
	\caption{\label{fig:impact_of_residual_stress}(color online).
		a) The probability distribution $P(s/\!\avg{s})$ of the residual shear stress divided by the ensemble average. $P(s/\!\avg{s})$ collapses onto a single curve and is independent of system size and pressure. Inset: $\avg{s} N^{1/2}$ as a function of pressure.
		b) $U_{AC}$ as a function of $pN^2$ for $N=256$ systems from the 2d $\Ec$ ensemble. Systems are binned according to the residual shear stress before averaging. For high $s$ (green diamonds), $U_{AC}$ is roughly constant but for low $s$ (blue squares), $U_{AC}$ is similar to the $\Eap$ data in Fig.~\ref{fig:anisotropic_AC_scaled_SS}.
		c) $D_{AC}$ displays the same behavior as $U_{AC}$.
		d) and e) Scatter plot of the lowest pressure values of $U_{AC}$ and $D_{AC}$, both of which show a remarkable linear dependence on the shear stress.
		The colors and symbols in a), d) and e) are the same as in Fig.~\ref{FigureRawData}.
	}
\end{figure}

\subsection{Anisotropy\label{mvh2}}

In this section we characterize the anisotropic modulations of the elastic constants. 

\subsubsection{Finite-size scaling of anisotropic elastic constant combinations}

As discussed above in Sec.~\ref{sec:usefulelasticconstants}, the elasticity of a jammed packing can be conveniently (though not completely
~\footnote{Since the full elasticity of an anisotropic system is described by 6 (21) independent constants in two (three) dimensions, the 5 quantities $B$, $G_{DC}$, $G_{AC}$, $U_{AC}$ and $D_{AC}$ are not sufficient to completely characterize a system's elastic properties. Unlike the elements of the elastic modulus tensor, however, they provide an intuitive description that conveniently isolates anisotropic fluctuations.}) described by the five quantities $B$, $G_{DC}$, $G_{AC}$, $U_{AC}$ and $D_{AC}$. The first two of these represent the average response to compression and shear, while the final three represent anisotropic fluctuations. Since anisotropy in jamming is a finite-size effect, one would expect the three ``$AC$" values to vanish in the thermodynamic limit. Here we explore their nontrivial dependence on system size and pressure, {\it i.e.}, proximity to the jamming transition. 
\begin{figure*}
	\centering
	\epsfig{file=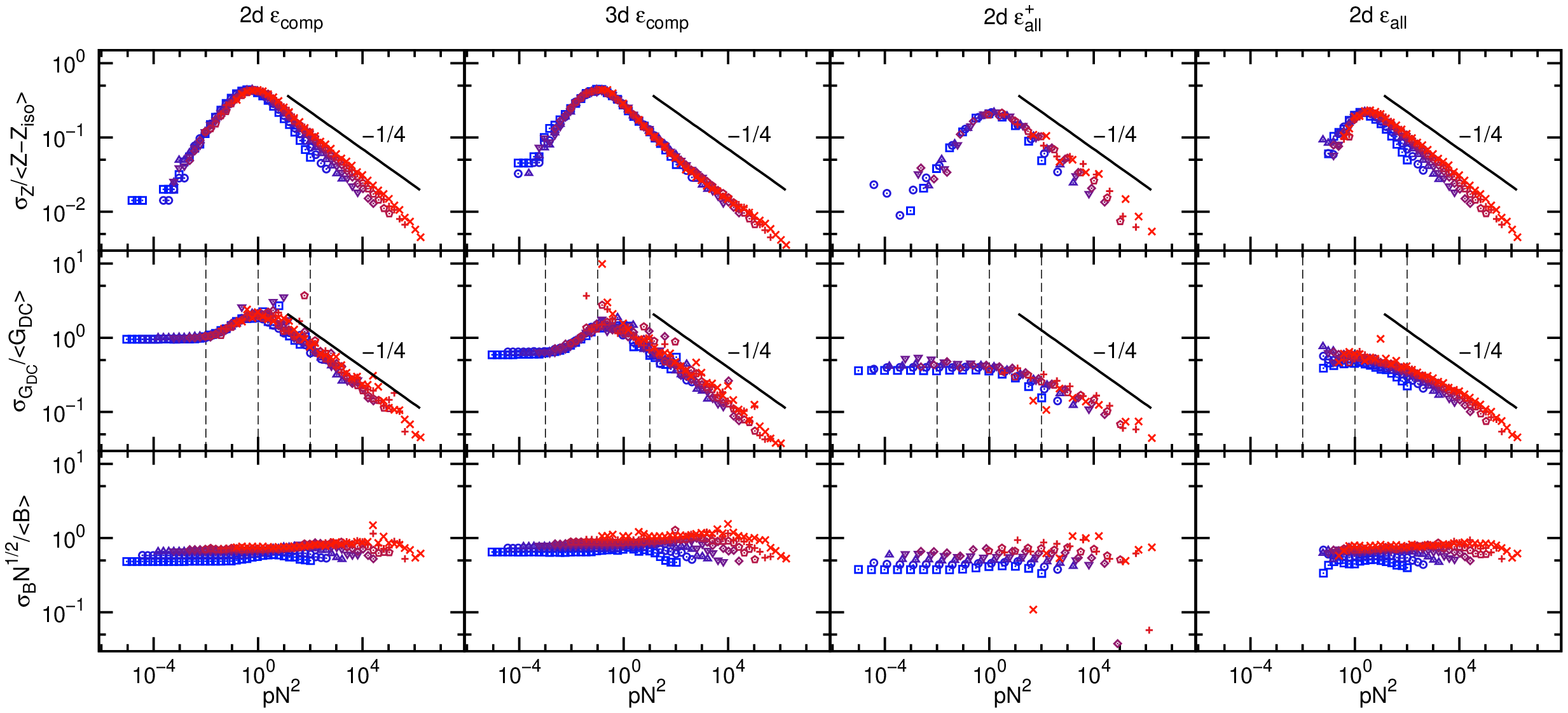,height=1\linewidth,angle=-90}
	\caption{\label{fig:anisotrpoic_sigma_scaled}(color online).
	The relative fluctuations in $Z$, $\Gdc$ and $B$. The colors and symbols are the same as in Fig.~\ref{FigureRawData}.}
\end{figure*}

\begin{figure*}
    \centering
    \begin{tabular}{c}
    	\epsfig{file=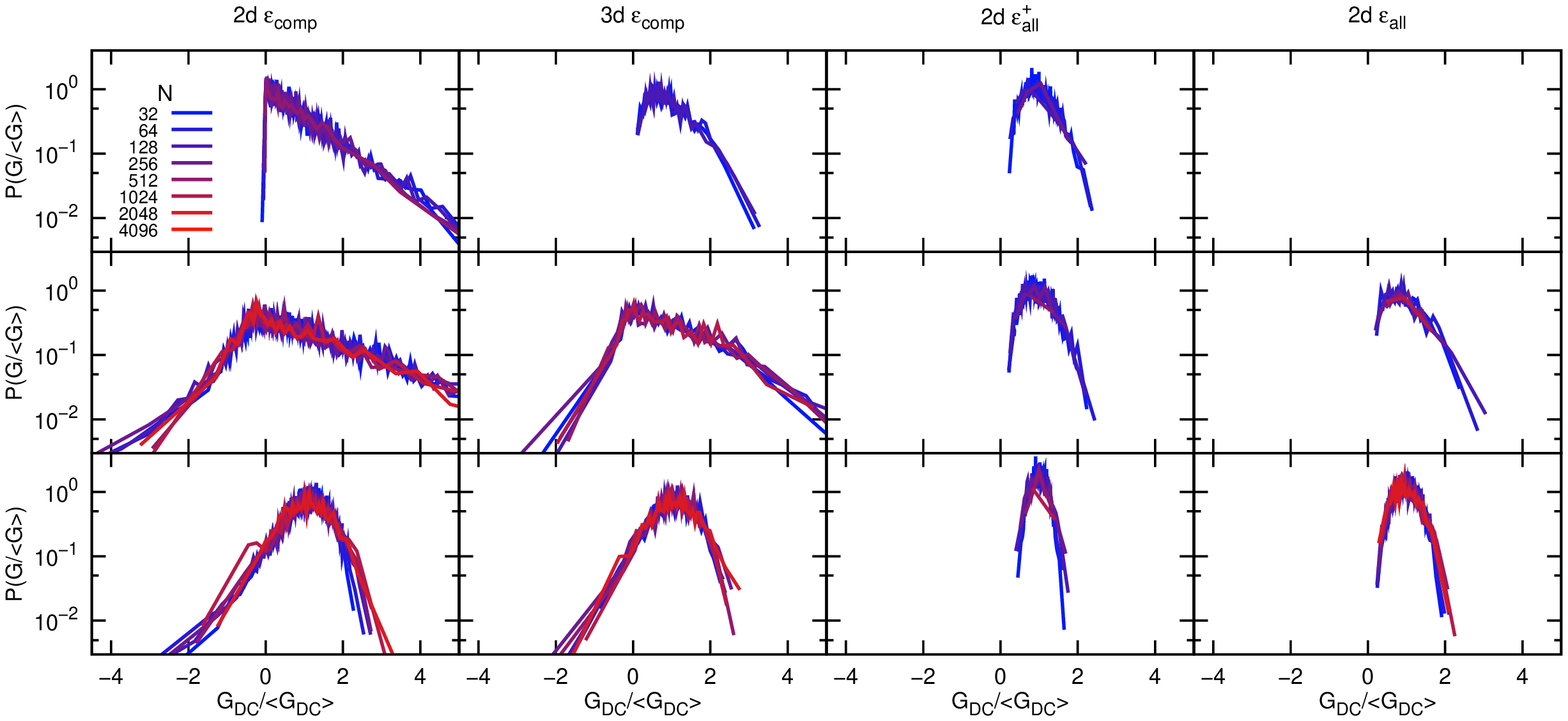,width=0.7\linewidth,angle=-90}
    \end{tabular}
    \caption{\label{fig:Gdist}(color online).
    The distribution of $G_{DC}$ for the different ensembles. The shape of the distribution function is different for low $pN^2$ (top row), medium $pN^2$ (middle row) and high $pN^2$ (bottom row), but collapses for systems at similar $pN^2$. The precise values of $pN^2$ correspond to the vertical dashed lines in Fig.~\ref{fig:anisotrpoic_sigma_scaled} ($10^{-1}$, $10^0$ and $10^1$ for the $2d$ ensembles; $10^{-1.5}$, $10^{-0.5}$ and $10^{0.5}$ for the $3d$ ensemble). }
\end{figure*}

The top row in Fig.~\ref{fig:anisotropic_AC_scaled} shows the anisotropic fluctuations of the shear modulus, $G_{AC}$, normalized by the average $\Gdc$ for all four ensembles. 
When plotted as a function of $pN^2$, the data collapse nicely onto a single curve, consistent with the finite-size scaling of Sec.~\ref{sec:finite_size_scaling}. We can distinguish three regimes, depending on the magnitude of $pN^2 $.

{\em (i)  $pN^2 \ll 1$:} Close to jamming, both $\Gdc$ and $G_{AC}$ are constant in pressure. For the two-dimensional $\Ec$ ensemble, the ratio $G_{AC}/\Gdc$ is approximately $1/\sqrt{2}$ (see the black dashed line). 
To understand this, first note that $G_{AC}$ is proportional to the peak height of the sinusoidal function $G(\theta)$ (see Fig.~\ref{fig:G_theta}), and the minimum of $G(\theta)$ is $G_\text{min} = \Gdc - \sqrt{2}G_{AC}$. Also note that $G_\text{min}$ is bounded by $-p$ at low pressures because a negative response can only arise from the pre-stress between contacts~\cite{Wyart:2005jna}. For the 2d $\Ec$ ensemble, we find that $G_\text{min}$ does indeed vanish as $p\rightarrow 0$, implying that $G_{AC}/\Gdc \rightarrow 1/\sqrt{2}$.
The fact that $G(\theta)$ reaches 0 (instead of remaining positive for all $\theta$) indicates that low $pN^2$ packings in this ensemble are on the edge of stability.  We note that while $G(\theta)$ is non-negative on average, it can nevertheless be negative for individual configurations included in the ensemble, either over a range of $\theta$ or even for all $\theta$, as noted earlier in Ref.~\cite{DagoisBohy:2012dh}.

{\em (ii)  $pN^2 \approx 1$:} In the crossover regime,
the minimum of $G(\theta)$ becomes negative for $\Ec$ packings, which implies that $G_{AC}/\Gdc > 1/\sqrt{2}$, leading to the characteristic ``bump'' in the $\Ec$ curves. However, this cannot happen for $\Ea$ or $\Eap$ packings because $G(\theta)$ must always be positive, and this bump is clearly absent there. 

{\em (iii)  $pN^2 \gg 1$:} At large pressures and system sizes, our results are consistent with the scaling $G_{AC}/\Gdc \sim \left(pN^2\right)^{-1/4}$. The $N$ dependence of this scaling is what one would expect from the central-limit theorem: relative fluctuations should be proportional to $1/\sqrt{N}$. The origin of the $p^{1/4}$ pressure dependence is not {\it a priori} obvious, but does follow if one assumes finite-size scaling with $pN^2$. Thus, the combination of the collapse in all three regimes with the non-trivial pressure dependence is strong evidence that finite-size scaling at the jamming transition is not a coincidence. Just as it is for classical phase transitions, finite-size scaling is a fundamental feature of jamming. 

The second row of Fig.~\ref{fig:anisotropic_AC_scaled} shows $U_{AC}$, normalized by the average $U_{DC}N^{-1/2}$. $U_{DC}$ itself is not shown but is given by Eq.~\eqref{eq:Udc_from_B_and_G}, and is constant at low pressures. The bottom row of Fig.~\ref{fig:anisotropic_AC_scaled} shows $D_{AC}$, which is normalized only by $N^{-1/2}$ because $D_{DC}=0$. For the $\Ec$ and $\Ea$ ensembles, $U_{AC}$ and $D_{AC}$ are constant at low and intermediate pressures, and deviate slightly at large pressures. They are also both proportional to the square root of the system size, again consistent with the central-limit theorem. 

From the data presented in Fig.~\ref{fig:anisotropic_AC_scaled} it is not clear if $U_{AC}$ and $D_{AC}$ collapse (note that the abscissa on these plots is $p$, not $pN^2$).  As we show below, there is solid evidence that these quantities have no single parameter scaling in the $\Ec$ and $\Ea$ ensembles. For the $\Eap$ ensemble (third column of Fig.~\ref{fig:anisotropic_AC_scaled}), $U_{AC}$ and $U_{DC}$ are qualitatively different. Interestingly, $U_{AC}N$ and $D_{AC}N$ in the $\Eap$ ensemble behave similarly to  $G_{AC}N$; as shown in Fig.~\ref{fig:anisotropic_AC_scaled_SS}, they are constant at low $pN^2$ and are proportional to $(pN^2)^{1/4}$ at high $pN^2$.  In the $\Eap$ ensemble, there is therefore clear evidence that $U_{AC}$ and $D_{AC}$ scale as $N^{-1/2}$ in the large $pN^2$ limit, consistent with expectations from the central limit theorem.

The discrepancy between the $\Eap$ and the other ensembles is due to the presence of residual shear stress in $\Ec$ and $\Ea$ packings. Figure~\ref{fig:impact_of_residual_stress}a shows that the distribution $P(s/\!\avg{s})$, where $s$ is the residual shear stress and $\avg{s}$ is the ensemble average, is independent of pressure and system size. In the inset, we see that $\avg{s}$ is roughly constant in pressure and is proportional to $N^{-1/2}$. To see the effect of the residual stress on $U_{AC}$ and $D_{AC}$, we bin systems according to $s$ and recalculate the average AC values. The results, which are shown in Fig.~\ref{fig:impact_of_residual_stress}b-c, clearly demonstrate the effect of residual stress on the low $pN^2$ behavior. For low $s$, $U_{AC}$ and $D_{AC}$ are similar to the $\Eap$ results in Fig.~\ref{fig:anisotropic_AC_scaled_SS}, where $s=0$ exactly. However, for high $s$, $U_{AC}$ and $D_{AC}$ are roughly flat. When considered together, the large $s$ data dominates the average leading to the lack of collapse seen in Fig.~\ref{fig:anisotropic_AC_scaled}.

\subsection{Statistical fluctuations in $Z-\Ziso$, $B$ and $\Gdc$ \label{sec:statistical_fluctuations}} 

In addition to $G_{AC}$, $U_{AC}$ and $D_{AC}$,  anisotropy effects can also be characterized by the distributions of contact number, bulk modulus and shear modulus. The simplest way to characterize these distributions is by their standard deviation. However, since the average quantities themselves change by many orders of magnitude, we normalize the standard deviations by the mean.

We begin with the distribution of the average number of contacts. The top row of Fig.~\ref{fig:anisotrpoic_sigma_scaled} shows the standard deviation $\sigma_Z$ of this distribution, normalized by the average of $Z-\Ziso$, which collapses as a function of $pN^2$. In the high and low $pN^2$ limits, the width of the distribution vanishes relative to the average. At intermediate $pN^2$, however, $\sigma_Z$ is of order $Z-\Ziso$. The second row of Fig.~\ref{fig:anisotrpoic_sigma_scaled} shows $\sigma_{\Gdc}$, which is almost identical to $G_{AC}$ (top row of Fig.~\ref{fig:anisotropic_AC_scaled}). Similarly, $\sigma_B$ is shown in the bottom row of Fig.~\ref{fig:anisotrpoic_sigma_scaled}. Interestingly, $\sigma_B$ is qualitatively similar to the high $s$ data for $U_{AC}$ in the $\Ec$ ensembles: $\sigma_B/B$ is proportional to $N^{-1/2}$ but roughly independent of pressure. The distinctive behavior of $U_{AC}$ in the $\Eap$ ensemble is not observed in $\sigma_B$.

One can also look at the full distributions of these quantities. We will focus on the shear modulus $\Gdc$. Fig.~\ref{fig:Gdist} shows the distribution of $\Gdc$, normalized by the average, for the four ensembles. The top, middle and bottom rows correspond to systems with low, intermediate and high $pN^2$, respectively, the precise values of which are given in the caption and depicted by vertical dashed lines in Fig.~\ref{fig:anisotrpoic_sigma_scaled}.

For a given ensemble, both the average of $\Gdc$ and $\sigma_{\Gdc}$ are independent of system size provided that $pN^2$ is held constant. Fig.~\ref{fig:Gdist} shows that this is true for the entire distribution of $\Gdc$. Indeed, the distribution can be considered a one parameter family of functions. Note that at low $pN^2$ (top row), the distribution vanishes very close to $\Gdc=0$ because, as discussed above, negative responses can only arise from stresses, which vanish with pressure.  At higher $pN^2$, however, $G_{DC}$ can be negative for the $\Ec$ ensemble.

\appendix

\section{Numerical Procedures\label{sec:numerical_procedures}}
A $d$ dimensional packing of $N$ spheres with equal mass $M$ is described by the position vectors $\vec r_{m}$ and radii $R_m$. Here, the index $m$ goes over the $N$ particles.
We will consider a simulation box with periodic boundaries made from the lattice vectors $\vec L_i$, where $i$ again indicates the dimension.
The center-center distance between particles $m$ and $m\p$ is given by
\eq{	r_{mm\p} = |\vec r_m-\vec r_{m\p} + \sum_b n_{mm\p}^i \vec L_i|,	} 
where $n_{mm\p}^i \in \{-1,0,1\}$ accounts for interactions across the periodic boundaries. 
The spheres interact via the harmonic soft-sphere potential
\eq{	U_{mm\p} = \frac \varepsilon 2 \left(1 - \frac{r_{mm\p}}{R_m+R_{m\p}}\right)^2	}
only when they overlap, {\it i.e.} when $r_{mm\p} < R_m + R_{m\p}$.
The units of length, mass, and energy are $D_\text{avg}$, $M$, and $\varepsilon$ respectively, where $D_\text{avg}\equiv N^{-1}\sum_m 2R_m$ is the average particle diameter.

\subsection{Generating sphere packings in the $\Ec$ ensemble}
To generate packings that satisfy the $\Rc$ requirement, we fix the lattice vectors:
\eq{	\vec L_i = L \vec e_i,}
where $\vec e_i$ is the unit vector in the $i$th direction. In other words, we use a standard cubic simulation box whose length $L$ is determined by the packing fraction $\phi$. 

In two dimensions, we choose the particles' radii to be uniformly distributed between 1 and 1.4 to prevent the issue discussed in Sec.~\ref{sec:JammingCriteria}I. In three dimensions, we use a 50/50 bidisperse mixture with ratio 1.4.
We begin by placing the particles at random at a very high packing fraction. We then quench the system to a zero temperature configuration by minimizing the total energy. We do this with a combination of line-search methods (L-BFGS and the Pollak-Ribi\`ere variant of Conjugate Gradient), the Newton-Rhapson method~\footnote{When calculating the inverse of the Hessian matrix in the Newton-Raphson method, we add to it $\lambda_0 \v I$, with $\v I$ the identity and $\lambda_0$ small, to suppress the global translations.}, and the FIRE algorithm~\cite{Bitzek:2006bw}. This combination of minimization algorithms was chosen to maximize accuracy and efficiency. However, given its speed, ease of implementation, and sensitivity to shallow features in the energy landscape, we would now recommend the exclusive use of the FIRE algorithm.

We then incrementally adjust the packing fraction, minimizing the energy after each iteration, until we are within $1\%$ of a desired pressure $p_\text{target}^1 = 10^{-1}$. Starting now with this configuration, we repeat this process with a slightly lower target pressure, $p_\text{target}^2 = 10^{-1.2}$. We continue lowering the target pressure incrementally until we reach $p_\text{target}^{36} = 10^{-8}$. Thus, for each initial random configuration, we obtain 36 states at logrithmically spaced pressures.

For each system size and dimension, we repeat this process for at least 1000 different initial random configurations. For small $N$ in two dimensions, we generate up to 5000 configurations to improve statistics. We do not consider systems for which the minimization algorithms fails to converge. This gives us the full two and three dimensional $\Ec$ ensembles. Finally, we can consider only the subset of systems that satisfy the $\Ra$ requirement to form the $\Ea$ ensemble.

\subsection{Generating sphere packings in the $\Eap$ ensemble}
To generate two dimensional packings that satisfy the $\Rap$ requirement, we also let the lattice vectors $\vec L_i$ vary. To separate the total volume from the shear degrees of freedom (and to suppress global rotations), we make the following change of variables:
\eqs{
	\vec{L}_1 &= L\left(\frac{1}{1+b},0\right) \\
	\vec{L}_2 &= L(a, 1+b).
}
The degrees of freedom of the system are thus the $dN$ components of the particle positions as well as $L$, $a$, and $b$. 
We then minimize the enthalpy-like potential introduced in Sec.~\ref{sec:precise_formulation}, 
\eq{H = U + p_\text{target} L^2, \label{eq:H_target_pressure}}
with respect to these $dN+3$ degrees of freedom.
This produces a system that 1) satisfies force balance at each particle, 2) has no residual shear stress, and 3) is at a pressure given precisely by $p_\text{target}$~\cite{DagoisBohy:2012dh}. 

Since minimizing Eq.~\eqref{eq:H_target_pressure} brings the system directly to the target pressure, we do not need to adjust the packing fraction manually. We also only use the Conjugate Gradient and FIRE~\cite{Bitzek:2006bw} algorithms. Note that in the FIRE algorithm, we set the effective mass of the boundary degrees of freedom to be $\sqrt{N}$.

\section{Elastic constants in two and three dimensions \label{AppendixA}}
Consider the symmetric, two dimensional strain tensor
\eq{ \overleftrightarrow{\epsilon}	 = \left( \begin{array}{cc} \epsilon_{xx} & \epsilon_{xy} \\  \epsilon_{xy}  &  \epsilon_{yy}  \end{array} \right).	}
We will consider the three dimensional case below.
This deformation is imposed on the system in accordance with Eq.~\eqref{strain_tensor_transformation}. After the system is allowed to relax, we define the response to be $R\equiv 2\frac{\Delta U}{V^0}$, where $\Delta U$ is the change in energy of the system and $V^0$ is the volume. To linear order, this is given in terms of the elastic modulus tensor:
\eqs{
	R = {}& c_{ijkl}\epsilon_{ij}\epsilon_{kl} \\
	= {} & c_{xxxx} \epsilon_{xx} ^2 + c_{yyyy} \epsilon_{yy}^2 \\
	&+ 4c_{xyxy} \epsilon_{xy}^2 + 2c_{xxyy}\epsilon_{xx}\epsilon_{yy} \\
	&+ 4c_{xxxy}\epsilon_{xx}\epsilon_{xy}+ 4c_{yyxy}\epsilon_{yy}\epsilon_{xy}. \label{R_from_cijkl}
}
Thus, if the 6 elastic constants $c_{xxxx}$, $c_{yyyy}$, $c_{xyxy}$, $c_{xxyy}$, $c_{xxxy}$, and $c_{yyxy}$ are known, then the linear response to any small deformation is easily obtained.

Although we are assuming that the system is not isotropic, there is no fundamental difference between the various directions -- the choice of axes is arbitrary. 
For a particular strain tensor, we can rotate the deformation by an angle $\theta$:
\eq{
	\overleftrightarrow{\epsilon}(\theta) &=
	\left( \begin{array}{cc} \cos\theta & \sin\theta \\ -\sin\theta & \cos\theta \end{array} \right)
	\left( \begin{array}{cc} \epsilon_{xx} & \epsilon_{xy} \\ \epsilon_{xy} & \epsilon_{yy} \end{array} \right)
	\left( \begin{array}{cc} \cos\theta & -\sin\theta \\ \sin\theta & \cos\theta \end{array} \right). \nonumber 
}
This results in a new deformation with a response $R(\theta)$.
Using the components of the rotated strain tensor,
\eq{	
	\epsilon_{xx}(\theta) &= \tfrac 1 2 (\epsilon_{xx}+\epsilon_{yy}) + \tfrac 1 2(\epsilon_{xx}-\epsilon_{yy})\cos2\theta + \epsilon_{xy} \sin2\theta \nonumber \\
	\epsilon_{yy}(\theta) &= \tfrac 1 2 (\epsilon_{xx}+\epsilon_{yy}) -   \tfrac 1 2(\epsilon_{xx}-\epsilon_{yy})\cos2\theta - \epsilon_{xy}\sin2\theta \nonumber \\
	\epsilon_{xy}(\theta) &=-\tfrac 1 2 (\epsilon_{xx}-\epsilon_{yy})\sin2\theta + \epsilon_{xy}\cos2\theta. \nonumber
}
the new response can be calculated from Eq.~\eqref{R_from_cijkl}.
Note that given the symmetry of Eq.~\eqref{R_from_cijkl}, $\theta$ can always be taken to be in the interval $[0,\pi]$.

By considering deformations that are rotations of each other, $R(\theta)$ is a convenient way to observe anisotropic fluctuations -- in an isotropic system, $R(\theta)$ is always independent of $\theta$. 
The first quantity of interest is the average response,
\eqs{
	R_{DC} &\equiv \avg{R(\theta)} \\
	& = \frac 1 {\pi}\int_0^{\pi}d\theta R(\theta), \label{Rbar}
}
which integrates out the anisotropic fluctuations. We can then characterize the anisotropy by the variance of the response:
\eqs{
	R_{AC}^2 &\equiv \avg{(R(\theta)- R_{DC})^2} \\
	&= \frac 1 {\pi} \int_0^{\pi}d\theta (R(\theta)- R_{DC})^2.  \label{sigmaR2}
}

Eqs.~\eqref{Rbar} and \eqref{sigmaR2} are generic in that we have not yet specified the initial strain tensor. Our strategy going forward will be to choose physically relevant strain tensors, {\it e.g.} corresponding to pure shear, calculate the response as a function of $\theta$, and use Eqs.~\eqref{Rbar} and \eqref{sigmaR2} to characterize the mean response as well as the fluctuations.
In doing so, it will be convenient to make the following definitions:
\eqs{
	G_0  &=c_{xyxy},\\
	G_{\tfrac{\pi}{4}} & = \frac 14 \left(c_{xxxx}+c_{yyyy} - 2c_{xxyy}\right) \\
	A_2 & = \sqrt{\frac 14 \left(c_{xxxx}-c_{yyyy}\right)^2 + \left(c_{xxxy}+c_{yyxy}\right)^2} \\
	\phi_2 &= \tan^{-1}\left(-2\left(c_{xxxy}+c_{yyxy}\right),c_{xxxx}-c_{yyyy}\right) \\
	A_4 &= -\frac 12 \sqrt{\left(c_{xxxy}-c_{yyxy}\right)^2 + \left(G_0 - G_{\tfrac{\pi}{4}}\right)^2} \\
	\phi_4 &= \tan^{-1}\left(c_{xxxy}-c_{yyxy},G_0 - G_{\tfrac{\pi}{4}}\right).
}

\subsection{Uniform Compression}
Uniform compression is obtained from the strain tensor
\eq{ \overleftrightarrow{\epsilon}	 = \frac \gamma 2 \left( \begin{array}{cc} 1 & 0 \\ 0 & 1 \end{array} \right),	}
 where we are interested in the limit $\gamma \ll 1$. This does not change under rotation and so the response, {\it i.e.} the bulk modulus $B$, can be calculated directly from Eq.~\eqref{R_from_cijkl}:
 \eq{	B = \frac 14 \left(c_{xxxx} + c_{yyyy} + 2c_{xxyy}\right).}

\subsection{Shear}
Pure shear can be obtained by setting $\epsilon_{xx} = \epsilon_{yy} = 0$ and $\epsilon_{xy} = \gamma/2$, resulting in the strain tensor
\eq{	\overleftrightarrow{\epsilon}(\theta) = \frac \gamma 2 \left( \begin{array}{cc} \sin(2\theta) & \cos(2\theta) \\ \cos(2\theta) & -\sin(2\theta) \end{array} \right),	}
where $\theta$ is the angle of shear. We will define $G(\theta)$ to be the response, which can be written as (see Fig.~\ref{fig:G_theta})
\eq{	G(\theta) = \frac 12 \left( G_0 + G_{\tfrac{\pi}{4}}  \right) - A_4 \sin \left( 4\theta + \phi_4 \right).	}
Note that although the generic period of $R(\theta)$ is $\pi$, $G(\theta)$ is periodic over the interval $[0,\pi/2]$.
Note also that $G(0)=G_0$ and $G(\pi/4) = G_{\tfrac{\pi}{4}}$.
From Eqs.~\eqref{Rbar} and \eqref{sigmaR2}, we see that
\eqs{
	G_{DC} &= \frac 12 \left( G_0 + G_{\tfrac{\pi}{4}}  \right) \\
	G_{AC} &= \frac {A_4}{\sqrt{2}}.
}

\subsection{Uniaxial Compression}
Uniaxial compression can be obtained by setting $\epsilon_{xx} = \gamma$ and $\epsilon_{yy} = \epsilon_{xy} = 0$, resulting in the strain tensor
\eq{	\overleftrightarrow{\epsilon}(\theta) = \frac \gamma 2 \left( \begin{array}{cc} 1+\cos(2\theta) & -\sin(2\theta) \\ -\sin(2\theta) & 1-\cos(2\theta) \end{array} \right).	}
We will define $U(\theta)$ to be the response, which can be written as
\eqs{	
	U(\theta) ={}& B + G_{DC} \\
	& + A_2 \sin(2\theta + \phi_2) \\
	&+ A_4 \sin(4\theta + \phi_4).
}
Note that $U(0) = c_{xxxx}$ and $U(\pi/2) = c_{yyyy}$.
From Eqs.~\eqref{Rbar} and \eqref{sigmaR2}, we see that
\eqs{
	U_{DC} &= B + G_{DC} \\
	U_{AC} &= \sqrt{ \frac 12 \left( A_2^2 + A_4^2 \right)}.
}

\subsection{Dilatancy}
Linear dilatancy can be understood from setting $\epsilon_{xx} = \epsilon_{xy} = \gamma/2$ and $\epsilon_{yy} = 0$, resulting in the strain tensor
\eq{	\epsilon(\theta) = \frac \gamma2 \left(
\begin{array}{cc}
 1+ \cos (2 \theta)+2\sin (2 \theta) & 2\cos (2 \theta)- \sin (2 \theta) \\
 2\cos (2 \theta)- \sin (2 \theta) & 1-\cos (2 \theta)-2\sin (2 \theta)
\end{array}
\right).
}
If the response of such a deformation is $R(\theta)$, then the dilatent response is
\eqs{	
	D(\theta) &= R(\theta) - \frac 14 U(\theta) - G(\theta) \\
	& = -\frac {A_2}2 \cos\left(2\theta + \phi_2\right) - A_4 \cos(4\theta + \phi_4).
}
When $\theta=0$, for example, we have from Eq.~\eqref{R_from_cijkl} that
\eq{	
	R(0) &= \frac 14 c_{xxxx} + c_{xyxy} + c_{xxxy} \\
	&= \frac 14 U(0) + G(0) + c_{xxxy}
}
so
\eq{	D(0) &= R(0) - \frac 14 U(0) - G(0) = c_{xxxy}.	}
Similarly, $D(\pi/2) = -c_{yyxy}$.
From Eqs.~\eqref{Rbar} and \eqref{sigmaR2}, we see that
\eqs{
	D_{DC} & = 0\\
	D_{AC} & = \sqrt{\frac 18 \left( A_2^2 + 4A_4^2 \right)}.
}
\\\\
\subsection{Three dimensions}
Extending the above definitions to three dimensions is straight forward. We begin with the strain tensor
\eq{	\overleftrightarrow{\epsilon} = \left(\begin{array}{ccc}\epsilon_{xx} & \epsilon_{xy} & \epsilon_{xz} \\ \epsilon_{xy} & \epsilon_{yy} & \epsilon_{yz} \\\epsilon_{xz} & \epsilon_{yz} & \epsilon_{zz}\end{array}\right) }
and the rotation matrix
\eq{	\mathcal{R}(\theta_1, \theta_2, \theta_3) = \mathcal{R}^\#(\theta_3) \cdot \mathcal{R}^*(\theta_2) \cdot \mathcal{R}^\#(\theta_1)}
where $\theta_1$, $\theta_2$ and $\theta_3$ are Euler angles and $\mathcal{R}^\#$ and $\mathcal{R}^*$ are given by
\eq{
	\mathcal{R}^\#(\theta) &=
		\left(	\begin{array}{ccc}
 			\cos \theta & -\sin \theta & 0 \\
			\sin \theta & \cos \theta & 0 \\
 			0 & 0 & 1
		\end{array} \right), \\
	\mathcal{R}^*(\theta) &=
		\left(	\begin{array}{ccc}
 			\cos \theta & 0 & \sin \theta \\
			0 & 1 & 0 \\
			-\sin \theta & 0 & \cos \theta
		\end{array} \right).
}
The rotated strain tensor,
\eq{	\overleftrightarrow{\epsilon} (\theta_1, \theta_2, \theta_3) = \mathcal{R}^{-1}(\theta_1, \theta_2, \theta_3) \cdot \overleftrightarrow{\epsilon} \cdot \mathcal{R}(\theta_1, \theta_2, \theta_3),}
and the response, $R(\theta_1, \theta_2, \theta_3)$, is a function of the three Euler angles.
Finally, the average response $R_{DC}$ and variance $R_{AC}^2$ are obtained from properly integrating over the three angles:
\eq{	
	R_{DC} &= \mathcal{I}^3\; R(\theta_1, \theta_2, \theta_3), \\
	R_{AC}^2 &= \mathcal{I}^3 \left[ R(\theta_1, \theta_2, \theta_3) - R_{DC} \right]^2,
}
where $\mathcal{I}^3$ stands for $ \frac{1}{32\pi^2}\int_0^{4\pi} d \theta_3 \int_0^\pi d \theta_2  \sin \theta_2 \int_0^{4\pi}  d \theta_1$.

\begin{acknowledgments}
We thank Wouter Ellenbroek, Silke Henkes, Tom Lubensky, Vincenzo Vitelli and Zorana Zeravcic for helpful discussions.
This research was supported by the U.S. Department of Energy, Office of Basic Energy Sciences, Division of Materials Sciences and Engineering under Awards DE-FG02-05ER46199 (A.J.L., C.P.G.) and DE-FG02-03ER46088 (S.R.N.). S.D.B. acknowledges funding from the Dutch physics foundation FOM, and B.P.T. and M.v.H. acknowledge funding from the Netherlands Organization for Scientific Research (NWO). C.P.G. was partially supported by the NSF through a Graduate Research Fellowship.
\end{acknowledgments}

% Create the reference section using BibTeX:
%\bibliography{papers2_bibtex,additional_bibtex}

\begin{thebibliography}{43}%
\makeatletter
\providecommand \@ifxundefined [1]{%
 \@ifx{#1\undefined}
}%
\providecommand \@ifnum [1]{%
 \ifnum #1\expandafter \@firstoftwo
 \else \expandafter \@secondoftwo
 \fi
}%
\providecommand \@ifx [1]{%
 \ifx #1\expandafter \@firstoftwo
 \else \expandafter \@secondoftwo
 \fi
}%
\providecommand \natexlab [1]{#1}%
\providecommand \enquote  [1]{``#1''}%
\providecommand \bibnamefont  [1]{#1}%
\providecommand \bibfnamefont [1]{#1}%
\providecommand \citenamefont [1]{#1}%
\providecommand \href@noop [0]{\@secondoftwo}%
\providecommand \href [0]{\begingroup \@sanitize@url \@href}%
\providecommand \@href[1]{\@@startlink{#1}\@@href}%
\providecommand \@@href[1]{\endgroup#1\@@endlink}%
\providecommand \@sanitize@url [0]{\catcode `\\12\catcode `\$12\catcode
  `\&12\catcode `\#12\catcode `\^12\catcode `\_12\catcode `\%12\relax}%
\providecommand \@@startlink[1]{}%
\providecommand \@@endlink[0]{}%
\providecommand \url  [0]{\begingroup\@sanitize@url \@url }%
\providecommand \@url [1]{\endgroup\@href {#1}{\urlprefix }}%
\providecommand \urlprefix  [0]{URL }%
\providecommand \Eprint [0]{\href }%
\providecommand \doibase [0]{http://dx.doi.org/}%
\providecommand \selectlanguage [0]{\@gobble}%
\providecommand \bibinfo  [0]{\@secondoftwo}%
\providecommand \bibfield  [0]{\@secondoftwo}%
\providecommand \translation [1]{[#1]}%
\providecommand \BibitemOpen [0]{}%
\providecommand \bibitemStop [0]{}%
\providecommand \bibitemNoStop [0]{.\EOS\space}%
\providecommand \EOS [0]{\spacefactor3000\relax}%
\providecommand \BibitemShut  [1]{\csname bibitem#1\endcsname}%
\let\auto@bib@innerbib\@empty
%</preamble>
\bibitem [{\citenamefont {Liu}\ and\ \citenamefont {Nagel}(2010)}]{Liu:2010jx}%
  \BibitemOpen
  \bibfield  {author} {\bibinfo {author} {\bibfnamefont {A.~J.}\ \bibnamefont
  {Liu}}\ and\ \bibinfo {author} {\bibfnamefont {S.~R.}\ \bibnamefont
  {Nagel}},\ }\href@noop {} {\bibfield  {journal} {\bibinfo  {journal} {Annu.
  Rev. Condens. Matter Phys.}\ }\textbf {\bibinfo {volume} {1}},\ \bibinfo
  {pages} {347} (\bibinfo {year} {2010})}\BibitemShut {NoStop}%
\bibitem [{\citenamefont {van Hecke}(2009)}]{vanHecke:2009go}%
  \BibitemOpen
  \bibfield  {author} {\bibinfo {author} {\bibfnamefont {M.}~\bibnamefont {van
  Hecke}},\ }\href@noop {} {\bibfield  {journal} {\bibinfo  {journal} {J.
  Phys.: Condens. Matter}\ }\textbf {\bibinfo {volume} {22}},\ \bibinfo {pages}
  {033101} (\bibinfo {year} {2009})}\BibitemShut {NoStop}%
\bibitem [{Note1()}]{Note1}%
  \BibitemOpen
  \bibinfo {note} {Note that the jamming transition appears to be a random
  first-order transition in dimensions $d \ge 2$, and is distinct from the
  glass transition, which is a random first-order transition in infinite
  dimensions~\cite {Parisi:2010uu}}\BibitemShut {NoStop}%
\bibitem [{\citenamefont {Durian}(1995)}]{Durian:1995eo}%
  \BibitemOpen
  \bibfield  {author} {\bibinfo {author} {\bibfnamefont {D.~J.}\ \bibnamefont
  {Durian}},\ }\href@noop {} {\bibfield  {journal} {\bibinfo  {journal} {Phys.
  Rev. Lett.}\ }\textbf {\bibinfo {volume} {75}},\ \bibinfo {pages} {4780}
  (\bibinfo {year} {1995})}\BibitemShut {NoStop}%
\bibitem [{\citenamefont {O'Hern}\ \emph {et~al.}(2003)\citenamefont {O'Hern},
  \citenamefont {Silbert}, \citenamefont {Liu},\ and\ \citenamefont
  {Nagel}}]{OHern:2003vq}%
  \BibitemOpen
  \bibfield  {author} {\bibinfo {author} {\bibfnamefont {C.~S.}\ \bibnamefont
  {O'Hern}}, \bibinfo {author} {\bibfnamefont {L.~E.}\ \bibnamefont {Silbert}},
  \bibinfo {author} {\bibfnamefont {A.~J.}\ \bibnamefont {Liu}}, \ and\
  \bibinfo {author} {\bibfnamefont {S.~R.}\ \bibnamefont {Nagel}},\ }\href@noop
  {} {\bibfield  {journal} {\bibinfo  {journal} {Phys. Rev. E}\ }\textbf
  {\bibinfo {volume} {68}},\ \bibinfo {pages} {011306} (\bibinfo {year}
  {2003})}\BibitemShut {NoStop}%
\bibitem [{\citenamefont {Silbert}\ \emph {et~al.}(2005)\citenamefont
  {Silbert}, \citenamefont {Liu},\ and\ \citenamefont
  {Nagel}}]{Silbert:2005vw}%
  \BibitemOpen
  \bibfield  {author} {\bibinfo {author} {\bibfnamefont {L.~E.}\ \bibnamefont
  {Silbert}}, \bibinfo {author} {\bibfnamefont {A.~J.}\ \bibnamefont {Liu}}, \
  and\ \bibinfo {author} {\bibfnamefont {S.~R.}\ \bibnamefont {Nagel}},\
  }\href@noop {} {\bibfield  {journal} {\bibinfo  {journal} {Phys. Rev. Lett.}\
  }\textbf {\bibinfo {volume} {95}},\ \bibinfo {pages} {098301} (\bibinfo
  {year} {2005})}\BibitemShut {NoStop}%
\bibitem [{\citenamefont {Wyart}(2005)}]{Wyart:2005vu}%
  \BibitemOpen
  \bibfield  {author} {\bibinfo {author} {\bibfnamefont {M.}~\bibnamefont
  {Wyart}},\ }\href@noop {} {\bibfield  {journal} {\bibinfo  {journal} {Ann
  Phys-Paris}\ }\textbf {\bibinfo {volume} {30}},\ \bibinfo {pages} {1}
  (\bibinfo {year} {2005})}\BibitemShut {NoStop}%
\bibitem [{\citenamefont {Wyart}\ \emph
  {et~al.}(2005{\natexlab{a}})\citenamefont {Wyart}, \citenamefont {Silbert},
  \citenamefont {Nagel},\ and\ \citenamefont {Witten}}]{Wyart:2005jna}%
  \BibitemOpen
  \bibfield  {author} {\bibinfo {author} {\bibfnamefont {M.}~\bibnamefont
  {Wyart}}, \bibinfo {author} {\bibfnamefont {L.~E.}\ \bibnamefont {Silbert}},
  \bibinfo {author} {\bibfnamefont {S.~R.}\ \bibnamefont {Nagel}}, \ and\
  \bibinfo {author} {\bibfnamefont {T.~A.}\ \bibnamefont {Witten}},\
  }\href@noop {} {\bibfield  {journal} {\bibinfo  {journal} {Phys. Rev. E}\
  }\textbf {\bibinfo {volume} {72}},\ \bibinfo {pages} {051306} (\bibinfo
  {year} {2005}{\natexlab{a}})}\BibitemShut {NoStop}%
\bibitem [{\citenamefont {Wyart}\ \emph
  {et~al.}(2005{\natexlab{b}})\citenamefont {Wyart}, \citenamefont {Nagel},\
  and\ \citenamefont {Witten}}]{Wyart:2005wv}%
  \BibitemOpen
  \bibfield  {author} {\bibinfo {author} {\bibfnamefont {M.}~\bibnamefont
  {Wyart}}, \bibinfo {author} {\bibfnamefont {S.~R.}\ \bibnamefont {Nagel}}, \
  and\ \bibinfo {author} {\bibfnamefont {T.~A.}\ \bibnamefont {Witten}},\
  }\href@noop {} {\bibfield  {journal} {\bibinfo  {journal} {EPL}\ }\textbf
  {\bibinfo {volume} {72}},\ \bibinfo {pages} {486} (\bibinfo {year}
  {2005}{\natexlab{b}})}\BibitemShut {NoStop}%
\bibitem [{\citenamefont {Silbert}\ \emph {et~al.}(2006)\citenamefont
  {Silbert}, \citenamefont {Liu},\ and\ \citenamefont
  {Nagel}}]{Silbert:2006bd}%
  \BibitemOpen
  \bibfield  {author} {\bibinfo {author} {\bibfnamefont {L.~E.}\ \bibnamefont
  {Silbert}}, \bibinfo {author} {\bibfnamefont {A.~J.}\ \bibnamefont {Liu}}, \
  and\ \bibinfo {author} {\bibfnamefont {S.~R.}\ \bibnamefont {Nagel}},\
  }\href@noop {} {\bibfield  {journal} {\bibinfo  {journal} {Phys. Rev. E}\
  }\textbf {\bibinfo {volume} {73}},\ \bibinfo {pages} {041304} (\bibinfo
  {year} {2006})}\BibitemShut {NoStop}%
\bibitem [{\citenamefont {Ellenbroek}\ \emph {et~al.}(2006)\citenamefont
  {Ellenbroek}, \citenamefont {Somfai}, \citenamefont {van Hecke},\ and\
  \citenamefont {van Saarloos}}]{Ellenbroek:2006df}%
  \BibitemOpen
  \bibfield  {author} {\bibinfo {author} {\bibfnamefont {W.~G.}\ \bibnamefont
  {Ellenbroek}}, \bibinfo {author} {\bibfnamefont {E.}~\bibnamefont {Somfai}},
  \bibinfo {author} {\bibfnamefont {M.}~\bibnamefont {van Hecke}}, \ and\
  \bibinfo {author} {\bibfnamefont {W.}~\bibnamefont {van Saarloos}},\
  }\href@noop {} {\bibfield  {journal} {\bibinfo  {journal} {Phys. Rev. Lett.}\
  }\textbf {\bibinfo {volume} {97}},\ \bibinfo {pages} {258001} (\bibinfo
  {year} {2006})}\BibitemShut {NoStop}%
\bibitem [{\citenamefont {Ellenbroek}\ \emph
  {et~al.}(2009{\natexlab{a}})\citenamefont {Ellenbroek}, \citenamefont {van
  Hecke},\ and\ \citenamefont {van Saarloos}}]{Ellenbroek:2009dp}%
  \BibitemOpen
  \bibfield  {author} {\bibinfo {author} {\bibfnamefont {W.~G.}\ \bibnamefont
  {Ellenbroek}}, \bibinfo {author} {\bibfnamefont {M.}~\bibnamefont {van
  Hecke}}, \ and\ \bibinfo {author} {\bibfnamefont {W.}~\bibnamefont {van
  Saarloos}},\ }\href@noop {} {\bibfield  {journal} {\bibinfo  {journal} {Phys.
  Rev. E}\ }\textbf {\bibinfo {volume} {80}},\ \bibinfo {pages} {061307}
  (\bibinfo {year} {2009}{\natexlab{a}})}\BibitemShut {NoStop}%
\bibitem [{\citenamefont {Goodrich}\ \emph {et~al.}(2013)\citenamefont
  {Goodrich}, \citenamefont {Ellenbroek},\ and\ \citenamefont
  {Liu}}]{Goodrich:2013ke}%
  \BibitemOpen
  \bibfield  {author} {\bibinfo {author} {\bibfnamefont {C.~P.}\ \bibnamefont
  {Goodrich}}, \bibinfo {author} {\bibfnamefont {W.~G.}\ \bibnamefont
  {Ellenbroek}}, \ and\ \bibinfo {author} {\bibfnamefont {A.~J.}\ \bibnamefont
  {Liu}},\ }\href@noop {} {\bibfield  {journal} {\bibinfo  {journal} {Soft
  Matter}\ }\textbf {\bibinfo {volume} {9}},\ \bibinfo {pages} {10993}
  (\bibinfo {year} {2013})}\BibitemShut {NoStop}%
\bibitem [{\citenamefont {Goodrich}\ \emph {et~al.}(2012)\citenamefont
  {Goodrich}, \citenamefont {Liu},\ and\ \citenamefont
  {Nagel}}]{Goodrich:2012ck}%
  \BibitemOpen
  \bibfield  {author} {\bibinfo {author} {\bibfnamefont {C.~P.}\ \bibnamefont
  {Goodrich}}, \bibinfo {author} {\bibfnamefont {A.~J.}\ \bibnamefont {Liu}}, \
  and\ \bibinfo {author} {\bibfnamefont {S.~R.}\ \bibnamefont {Nagel}},\
  }\href@noop {} {\bibfield  {journal} {\bibinfo  {journal} {Phys. Rev. Lett.}\
  }\textbf {\bibinfo {volume} {109}},\ \bibinfo {pages} {095704} (\bibinfo
  {year} {2012})}\BibitemShut {NoStop}%
\bibitem [{\citenamefont {Torquato}\ and\ \citenamefont
  {Stillinger}(2001)}]{Torquato:2001bm}%
  \BibitemOpen
  \bibfield  {author} {\bibinfo {author} {\bibfnamefont {S.}~\bibnamefont
  {Torquato}}\ and\ \bibinfo {author} {\bibfnamefont {F.~H.}\ \bibnamefont
  {Stillinger}},\ }\href@noop {} {\bibfield  {journal} {\bibinfo  {journal} {J.
  Phys. Chem. B}\ }\textbf {\bibinfo {volume} {105}},\ \bibinfo {pages} {11849}
  (\bibinfo {year} {2001})}\BibitemShut {NoStop}%
\bibitem [{\citenamefont {Dagois-Bohy}\ \emph {et~al.}(2012)\citenamefont
  {Dagois-Bohy}, \citenamefont {Tighe}, \citenamefont {Simon}, \citenamefont
  {Henkes},\ and\ \citenamefont {van Hecke}}]{DagoisBohy:2012dh}%
  \BibitemOpen
  \bibfield  {author} {\bibinfo {author} {\bibfnamefont {S.}~\bibnamefont
  {Dagois-Bohy}}, \bibinfo {author} {\bibfnamefont {B.~P.}~\bibnamefont {Tighe}},
  \bibinfo {author} {\bibfnamefont {J.}~\bibnamefont {Simon}}, \bibinfo
  {author} {\bibfnamefont {S.}~\bibnamefont {Henkes}}, \ and\ \bibinfo {author}
  {\bibfnamefont {M.}~\bibnamefont {van Hecke}},\ }\href@noop {} {\bibfield
  {journal} {\bibinfo  {journal} {Phys. Rev. Lett.}\ }\textbf {\bibinfo
  {volume} {109}},\ \bibinfo {pages} {095703} (\bibinfo {year}
  {2012})}\BibitemShut {NoStop}%
\bibitem [{\citenamefont {Charbonneau}\ \emph {et~al.}(2012)\citenamefont
  {Charbonneau}, \citenamefont {Corwin}, \citenamefont {Parisi},\ and\
  \citenamefont {Zamponi}}]{Charbonneau:2012fl}%
  \BibitemOpen
  \bibfield  {author} {\bibinfo {author} {\bibfnamefont {P.}~\bibnamefont
  {Charbonneau}}, \bibinfo {author} {\bibfnamefont {E.~I.}\ \bibnamefont
  {Corwin}}, \bibinfo {author} {\bibfnamefont {G.}~\bibnamefont {Parisi}}, \
  and\ \bibinfo {author} {\bibfnamefont {F.}~\bibnamefont {Zamponi}},\
  }\href@noop {} {\bibfield  {journal} {\bibinfo  {journal} {Phys. Rev. Lett.}\
  }\textbf {\bibinfo {volume} {109}},\ \bibinfo {pages} {205501} (\bibinfo
  {year} {2012})}\BibitemShut {NoStop}%
  %
\bibitem [{\citenamefont {V{\aa}gberg}\ \emph {et~al.}(2011)\citenamefont
  {V{\aa}gberg}, \citenamefont {Valdez-Balderas}, \citenamefont {Moore},
  \citenamefont {Olsson},\ and\ \citenamefont {Teitel}}]{Vagberg:2011fe}%
  \BibitemOpen
  \bibfield  {author} {\bibinfo {author} {\bibfnamefont {D.}~\bibnamefont
  {V{\aa}gberg}}, \bibinfo {author} {\bibfnamefont {D.}~\bibnamefont
  {Valdez-Balderas}}, \bibinfo {author} {\bibfnamefont {M.~A.}~\bibnamefont
  {Moore}}, \bibinfo {author} {\bibfnamefont {P.}~\bibnamefont {Olsson}}, \
  and\ \bibinfo {author} {\bibfnamefont {S.}~\bibnamefont {Teitel}},\
  }\href@noop {} {\bibfield  {journal} {\bibinfo  {journal} {Phys. Rev. E}\
  }\textbf {\bibinfo {volume} {83}},\ \bibinfo {pages} {030303} (\bibinfo
  {year} {2011})}\BibitemShut {NoStop}%
%
\bibitem [{\citenamefont {Chaudhuri}\ \emph {et~al.}(2010)\citenamefont
  {Chaudhuri}, \citenamefont {Berthier},\ and\ \citenamefont
  {Sastry}}]{Chaudhuri:2010jg}%
  \BibitemOpen
  \bibfield  {author} {\bibinfo {author} {\bibfnamefont {P.}~\bibnamefont
  {Chaudhuri}}, \bibinfo {author} {\bibfnamefont {L.}~\bibnamefont {Berthier}},
  \ and\ \bibinfo {author} {\bibfnamefont {S.}~\bibnamefont {Sastry}},\
  }\href@noop {} {\bibfield  {journal} {\bibinfo  {journal} {Phys. Rev. Lett.}\
  }\textbf {\bibinfo {volume} {104}},\ \bibinfo {pages} {165701} (\bibinfo
  {year} {2010})}\BibitemShut {NoStop}%
\bibitem [{\citenamefont {Liu}\ \emph {et~al.}(2014)\citenamefont {Liu},
  \citenamefont {Xie},\ and\ \citenamefont {Xu}}]{Liu:2014gu}%
  \BibitemOpen
  \bibfield  {author} {\bibinfo {author} {\bibfnamefont {H.}~\bibnamefont
  {Liu}}, \bibinfo {author} {\bibfnamefont {X.}~\bibnamefont {Xie}}, \ and\
  \bibinfo {author} {\bibfnamefont {N.}~\bibnamefont {Xu}},\ }\href@noop {}
  {\bibfield  {journal} {\bibinfo  {journal} {Phys. Rev. Lett.}\ }\textbf
  {\bibinfo {volume} {112}},\ \bibinfo {pages} {145502} (\bibinfo {year}
  {2014})}\BibitemShut {NoStop}%
\bibitem [{\citenamefont {Bolton}\ and\ \citenamefont
  {Weaire}(1990)}]{BOLTON:1990uy}%
  \BibitemOpen
  \bibfield  {author} {\bibinfo {author} {\bibfnamefont {F.}~\bibnamefont
  {Bolton}}\ and\ \bibinfo {author} {\bibfnamefont {D.}~\bibnamefont
  {Weaire}},\ }\href@noop {} {\bibfield  {journal} {\bibinfo  {journal} {Phys.
  Rev. Lett.}\ }\textbf {\bibinfo {volume} {65}},\ \bibinfo {pages} {3449}
  (\bibinfo {year} {1990})}\BibitemShut {NoStop}%
\bibitem [{\citenamefont {Alexander}(1998)}]{Alexander:1998vc}%
  \BibitemOpen
  \bibfield  {author} {\bibinfo {author} {\bibfnamefont {S.}~\bibnamefont
  {Alexander}},\ }\href@noop {} {\bibfield  {journal} {\bibinfo  {journal}
  {Physics Reports}\ }\textbf {\bibinfo {volume} {296}},\ \bibinfo {pages} {65}
  (\bibinfo {year} {1998})}\BibitemShut {NoStop}%
\bibitem [{\citenamefont {Moukarzel}(1998)}]{Moukarzel:1998vn}%
  \BibitemOpen
  \bibfield  {author} {\bibinfo {author} {\bibfnamefont {C.~F.}\ \bibnamefont
  {Moukarzel}},\ }\href@noop {} {\bibfield  {journal} {\bibinfo  {journal}
  {Phys. Rev. Lett.}\ }\textbf {\bibinfo {volume} {81}},\ \bibinfo {pages}
  {1634} (\bibinfo {year} {1998})}\BibitemShut {NoStop}%
\bibitem [{\citenamefont {Donev}\ \emph {et~al.}(2007)\citenamefont {Donev},
  \citenamefont {Connelly}, \citenamefont {Stillinger},\ and\ \citenamefont
  {Torquato}}]{Donev:2007go}%
  \BibitemOpen
  \bibfield  {author} {\bibinfo {author} {\bibfnamefont {A.}~\bibnamefont
  {Donev}}, \bibinfo {author} {\bibfnamefont {R.}~\bibnamefont {Connelly}},
  \bibinfo {author} {\bibfnamefont {F.~H.}\ \bibnamefont {Stillinger}}, \ and\
  \bibinfo {author} {\bibfnamefont {S.}~\bibnamefont {Torquato}},\ }\href@noop
  {} {\bibfield  {journal} {\bibinfo  {journal} {Phys. Rev. E}\ }\textbf
  {\bibinfo {volume} {75}},\ \bibinfo {pages} {051304} (\bibinfo {year}
  {2007})}\BibitemShut {NoStop}%
\bibitem [{\citenamefont {Zeravcic}\ \emph {et~al.}(2009)\citenamefont
  {Zeravcic}, \citenamefont {Xu}, \citenamefont {Liu}, \citenamefont {Nagel},\
  and\ \citenamefont {van Saarloos}}]{Zeravcic:2009wo}%
  \BibitemOpen
  \bibfield  {author} {\bibinfo {author} {\bibfnamefont {Z.}~\bibnamefont
  {Zeravcic}}, \bibinfo {author} {\bibfnamefont {N.}~\bibnamefont {Xu}},
  \bibinfo {author} {\bibfnamefont {A.~J.}\ \bibnamefont {Liu}}, \bibinfo
  {author} {\bibfnamefont {S.~R.}\ \bibnamefont {Nagel}}, \ and\ \bibinfo
  {author} {\bibfnamefont {W.}~\bibnamefont {van Saarloos}},\ }\href@noop {}
  {\bibfield  {journal} {\bibinfo  {journal} {EPL}\ }\textbf {\bibinfo {volume}
  {87}},\ \bibinfo {pages} {26001} (\bibinfo {year} {2009})}\BibitemShut
  {NoStop}%
\bibitem [{\citenamefont {Mailman}\ \emph {et~al.}(2009)\citenamefont
  {Mailman}, \citenamefont {Schreck}, \citenamefont {O'Hern},\ and\
  \citenamefont {Chakraborty}}]{Mailman:2009ct}%
  \BibitemOpen
  \bibfield  {author} {\bibinfo {author} {\bibfnamefont {M.}~\bibnamefont
  {Mailman}}, \bibinfo {author} {\bibfnamefont {C.~F.}\ \bibnamefont
  {Schreck}}, \bibinfo {author} {\bibfnamefont {C.~S.}\ \bibnamefont {O'Hern}},
  \ and\ \bibinfo {author} {\bibfnamefont {B.}~\bibnamefont {Chakraborty}},\
  }\href@noop {} {\bibfield  {journal} {\bibinfo  {journal} {Phys. Rev. Lett.}\
  }\textbf {\bibinfo {volume} {102}},\ \bibinfo {pages} {255501} (\bibinfo
  {year} {2009})}\BibitemShut {NoStop}%
\bibitem [{\citenamefont {Shundyak}\ \emph {et~al.}(2007)\citenamefont
  {Shundyak}, \citenamefont {van Hecke},\ and\ \citenamefont {van
  Saarloos}}]{Shundyak:2007ga}%
  \BibitemOpen
  \bibfield  {author} {\bibinfo {author} {\bibfnamefont {K.}~\bibnamefont
  {Shundyak}}, \bibinfo {author} {\bibfnamefont {M.}~\bibnamefont {van Hecke}},
  \ and\ \bibinfo {author} {\bibfnamefont {W.}~\bibnamefont {van Saarloos}},\
  }\href@noop {} {\bibfield  {journal} {\bibinfo  {journal} {Phys. Rev. E}\
  }\textbf {\bibinfo {volume} {75}},\ \bibinfo {pages} {010301} (\bibinfo
  {year} {2007})}\BibitemShut {NoStop}%
\bibitem [{\citenamefont {Somfai}\ \emph {et~al.}(2007)\citenamefont {Somfai},
  \citenamefont {van Hecke}, \citenamefont {Ellenbroek}, \citenamefont
  {Shundyak},\ and\ \citenamefont {van Saarloos}}]{Somfai:2007ge}%
  \BibitemOpen
  \bibfield  {author} {\bibinfo {author} {\bibfnamefont {E.}~\bibnamefont
  {Somfai}}, \bibinfo {author} {\bibfnamefont {M.}~\bibnamefont {van Hecke}},
  \bibinfo {author} {\bibfnamefont {W.~G.}\ \bibnamefont {Ellenbroek}},
  \bibinfo {author} {\bibfnamefont {K.}~\bibnamefont {Shundyak}}, \ and\
  \bibinfo {author} {\bibfnamefont {W.}~\bibnamefont {van Saarloos}},\
  }\href@noop {} {\bibfield  {journal} {\bibinfo  {journal} {Phys. Rev. E}\
  }\textbf {\bibinfo {volume} {75}},\ \bibinfo {pages} {020301} (\bibinfo
  {year} {2007})}\BibitemShut {NoStop}%
\bibitem [{\citenamefont {Henkes}\ \emph
  {et~al.}(2010{\natexlab{a}})\citenamefont {Henkes}, \citenamefont {Shundyak},
  \citenamefont {van Saarloos},\ and\ \citenamefont {van
  Hecke}}]{Henkes:2010uu}%
  \BibitemOpen
  \bibfield  {author} {\bibinfo {author} {\bibfnamefont {S.}~\bibnamefont
  {Henkes}}, \bibinfo {author} {\bibfnamefont {K.}~\bibnamefont {Shundyak}},
  \bibinfo {author} {\bibfnamefont {W.}~\bibnamefont {van Saarloos}}, \ and\
  \bibinfo {author} {\bibfnamefont {M.}~\bibnamefont {van Hecke}},\ }\href@noop
  {} {\bibfield  {journal} {\bibinfo  {journal} {Soft Matter}\ }\textbf
  {\bibinfo {volume} {6}},\ \bibinfo {pages} {2935} (\bibinfo {year}
  {2010}{\natexlab{a}})}\BibitemShut {NoStop}%
\bibitem [{\citenamefont {Henkes}\ \emph
  {et~al.}(2010{\natexlab{b}})\citenamefont {Henkes}, \citenamefont {van
  Hecke},\ and\ \citenamefont {van Saarloos}}]{Henkes:2010kv}%
  \BibitemOpen
  \bibfield  {author} {\bibinfo {author} {\bibfnamefont {S.}~\bibnamefont
  {Henkes}}, \bibinfo {author} {\bibfnamefont {M.}~\bibnamefont {van Hecke}}, \
  and\ \bibinfo {author} {\bibfnamefont {W.}~\bibnamefont {van Saarloos}},\
  }\href@noop {} {\bibfield  {journal} {\bibinfo  {journal} {EPL}\ }\textbf
  {\bibinfo {volume} {90}},\ \bibinfo {pages} {14003} (\bibinfo {year}
  {2010}{\natexlab{b}})}\BibitemShut {NoStop}%
\bibitem [{\citenamefont {Papanikolaou}\ \emph {et~al.}(2013)\citenamefont
  {Papanikolaou}, \citenamefont {O'Hern},\ and\ \citenamefont
  {Shattuck}}]{Papanikolaou:2013fa}%
  \BibitemOpen
  \bibfield  {author} {\bibinfo {author} {\bibfnamefont {S.}~\bibnamefont
  {Papanikolaou}}, \bibinfo {author} {\bibfnamefont {C.~S.}\ \bibnamefont
  {O'Hern}}, \ and\ \bibinfo {author} {\bibfnamefont {M.~D.}\ \bibnamefont
  {Shattuck}},\ }\href@noop {} {\bibfield  {journal} {\bibinfo  {journal}
  {Phys. Rev. Lett.}\ }\textbf {\bibinfo {volume} {110}},\ \bibinfo {pages}
  {198002} (\bibinfo {year} {2013})}\BibitemShut {NoStop}%
\bibitem [{Note2()}]{Note2}%
  \BibitemOpen
  \bibinfo {note} {We find the difference to be small, of order
  $10^{-3}$.}\BibitemShut {Stop}%
\bibitem [{\citenamefont {Torquato}\ and\ \citenamefont
  {Jiao}(2010)}]{Torquato:2010hb}%
  \BibitemOpen
  \bibfield  {author} {\bibinfo {author} {\bibfnamefont {S.}~\bibnamefont
  {Torquato}}\ and\ \bibinfo {author} {\bibfnamefont {Y.}~\bibnamefont
  {Jiao}},\ }\href@noop {} {\bibfield  {journal} {\bibinfo  {journal} {Phys.
  Rev. E (3)}\ }\textbf {\bibinfo {volume} {82}},\ \bibinfo {pages} {061302}
  (\bibinfo {year} {2010})}\BibitemShut {NoStop}%
\bibitem [{\citenamefont {Tighe}(2011)}]{Tighe:2011fq}%
  \BibitemOpen
  \bibfield  {author} {\bibinfo {author} {\bibfnamefont {B.~P.}~\bibnamefont
  {Tighe}},\ }\href@noop {} {\bibfield  {journal} {\bibinfo  {journal} {Phys.
  Rev. Lett.}\ }\textbf {\bibinfo {volume} {107}},\ \bibinfo {pages} {158303}
  (\bibinfo {year} {2011})}\BibitemShut {NoStop}%
\bibitem [{\citenamefont {Maloney}\ and\ \citenamefont
  {Lema{\^\i}tre}(2006)}]{Maloney:2006dt}%
  \BibitemOpen
  \bibfield  {author} {\bibinfo {author} {\bibfnamefont {C.~E.}~\bibnamefont
  {Maloney}}\ and\ \bibinfo {author} {\bibfnamefont {A.}~\bibnamefont
  {Lemaitre}},\ }\href@noop {} {\bibfield  {journal} {\bibinfo  {journal}
  {Phys. Rev. E}\ }\textbf {\bibinfo {volume} {74}},\ \bibinfo {pages} {016118}
  (\bibinfo {year} {2006})}\BibitemShut {NoStop}%
\bibitem [{\citenamefont {Stenull}\ and\ \citenamefont
  {Lubensky}(2014)}]{StenullLubensky}%
  \BibitemOpen
  \bibfield  {author} {\bibinfo {author} {\bibfnamefont {O.}~\bibnamefont
  {Stenull}}\ and\ \bibinfo {author} {\bibfnamefont {T.~C.}\ \bibnamefont
  {Lubensky}},\ }\href@noop {} {\bibfield  {journal} {\bibinfo  {journal} {in
  preparation}\ } (\bibinfo {year} {2014})}\BibitemShut {NoStop}%
\bibitem [{\citenamefont {Ellenbroek}\ \emph
  {et~al.}(2009{\natexlab{b}})\citenamefont {Ellenbroek}, \citenamefont
  {Zeravcic}, \citenamefont {van Saarloos},\ and\ \citenamefont {van
  Hecke}}]{Ellenbroek:2009to}%
  \BibitemOpen
  \bibfield  {author} {\bibinfo {author} {\bibfnamefont {W.~G.}\ \bibnamefont
  {Ellenbroek}}, \bibinfo {author} {\bibfnamefont {Z.}~\bibnamefont
  {Zeravcic}}, \bibinfo {author} {\bibfnamefont {W.}~\bibnamefont {van
  Saarloos}}, \ and\ \bibinfo {author} {\bibfnamefont {M.}~\bibnamefont {van
  Hecke}},\ }\href@noop {} {\bibfield  {journal} {\bibinfo  {journal} {EPL}\
  }\textbf {\bibinfo {volume} {87}},\ \bibinfo {pages} {34004} (\bibinfo {year}
  {2009}{\natexlab{b}})}\BibitemShut {NoStop}%
\bibitem [{\citenamefont {Binder}\ \emph {et~al.}(1985)\citenamefont {Binder},
  \citenamefont {Nauenberg}, \citenamefont {Privman},\ and\ \citenamefont
  {Young}}]{BINDER:1985vl}%
  \BibitemOpen
  \bibfield  {author} {\bibinfo {author} {\bibfnamefont {K.}~\bibnamefont
  {Binder}}, \bibinfo {author} {\bibfnamefont {M.}~\bibnamefont {Nauenberg}},
  \bibinfo {author} {\bibfnamefont {V.}~\bibnamefont {Privman}}, \ and\
  \bibinfo {author} {\bibfnamefont {A.~P.}~\bibnamefont {Young}},\ }\href@noop {}
  {\bibfield  {journal} {\bibinfo  {journal} {Phys. Rev. B}\ }\textbf {\bibinfo
  {volume} {31}},\ \bibinfo {pages} {1498} (\bibinfo {year}
  {1985})}\BibitemShut {NoStop}%
\bibitem [{\citenamefont {Dillmann}\ \emph {et~al.}(1998)\citenamefont
  {Dillmann}, \citenamefont {Janke},\ and\ \citenamefont
  {Binder}}]{Dillmann:1998ty}%
  \BibitemOpen
  \bibfield  {author} {\bibinfo {author} {\bibfnamefont {O.}~\bibnamefont
  {Dillmann}}, \bibinfo {author} {\bibfnamefont {W.}~\bibnamefont {Janke}}, \
  and\ \bibinfo {author} {\bibfnamefont {K.}~\bibnamefont {Binder}},\
  }\href@noop {} {\bibfield  {journal} {\bibinfo  {journal} {J Stat Phys}\
  }\textbf {\bibinfo {volume} {92}},\ \bibinfo {pages} {57} (\bibinfo {year}
  {1998})}\BibitemShut {NoStop}%
\bibitem [{Note3()}]{Note3}%
  \BibitemOpen
  \bibinfo {note} {Since the full elasticity of an anisotropic system is
  described by 6 (21) independent constants in two (three) dimensions, the 5
  quantities $B$, $G_{DC}$, $G_{AC}$, $U_{AC}$ and $D_{AC}$ are not sufficient
  to completely characterize a system's elastic properties. Unlike the elements
  of the elastic modulus tensor, however, they provide an intuitive description
  that conveniently isolates anisotropic fluctuations.}\BibitemShut {Stop}%
\bibitem [{Note4()}]{Note4}%
  \BibitemOpen
  \bibinfo {note} {When calculating the inverse of the Hessian matrix in the
  Newton-Raphson method, we add to it $\lambda _0 \protect \ensuremath
  {\protect \mathbf {I}}$, with $\protect \ensuremath {\protect \mathbf {I}}$
  the identity and $\lambda _0$ small, to suppress the global
  translations.}\BibitemShut {Stop}%
\bibitem [{\citenamefont {Bitzek}\ \emph {et~al.}(2006)\citenamefont {Bitzek},
  \citenamefont {Koskinen}, \citenamefont {G{\"a}hler}, \citenamefont
  {Moseler},\ and\ \citenamefont {Gumbsch}}]{Bitzek:2006bw}%
  \BibitemOpen
  \bibfield  {author} {\bibinfo {author} {\bibfnamefont {E.}~\bibnamefont
  {Bitzek}}, \bibinfo {author} {\bibfnamefont {P.}~\bibnamefont {Koskinen}},
  \bibinfo {author} {\bibfnamefont {F.}~\bibnamefont {G{\"a}hler}}, \bibinfo
  {author} {\bibfnamefont {M.}~\bibnamefont {Moseler}}, \ and\ \bibinfo
  {author} {\bibfnamefont {P.}~\bibnamefont {Gumbsch}},\ }\href@noop {}
  {\bibfield  {journal} {\bibinfo  {journal} {Phys. Rev. Lett.}\ }\textbf
  {\bibinfo {volume} {97}},\ \bibinfo {pages} {170201} (\bibinfo {year}
  {2006})}\BibitemShut {NoStop}%
\bibitem [{\citenamefont {Parisi}\ and\ \citenamefont
  {Zamponi}(2010)}]{Parisi:2010uu}%
  \BibitemOpen
  \bibfield  {author} {\bibinfo {author} {\bibfnamefont {G.}~\bibnamefont
  {Parisi}}\ and\ \bibinfo {author} {\bibfnamefont {F.}~\bibnamefont
  {Zamponi}},\ }\href@noop {} {\bibfield  {journal} {\bibinfo  {journal}
  {Reviews of Modern Physics}\ }\textbf {\bibinfo {volume} {82}},\ \bibinfo
  {pages} {789} (\bibinfo {year} {2010})}\BibitemShut {NoStop}%
\end{thebibliography}

%merlin.mbs apsrev4-1.bst 2010-07-25 4.21a (PWD, AO, DPC) hacked
%Control: key (0)
%Control: author (8) initials jnrlst
%Control: editor formatted (1) identically to author
%Control: production of article title (-1) disabled
%Control: page (0) single
%Control: year (1) truncated
%Control: production of eprint (0) enabled
%

\end{document}